\documentclass[a4paper,fleqn,usenatbib,useAMS]{mnras}
\usepackage{graphics}
\usepackage{graphicx}
\usepackage{amsmath}    
\usepackage{amssymb}    
\usepackage{multicol}        
\usepackage{bm}     
\usepackage{pdflscape}  
\usepackage{xcolor}
\usepackage{ae,aecompl}
\usepackage{subfigure}
\usepackage{scalerel}
\usepackage[english]{babel}
\usepackage{times}
\usepackage{physics}
\usepackage{etoolbox}
\usepackage{paralist}
\usepackage[inline]{enumitem}
\newlist{mycompactenum}{enumerate}{1}
\setlist[mycompactenum,1]{nosep,label=\arabic*.}
\usepackage{paralist}
\makeatletter
\patchcmd\@combinedblfloats{\box\@outputbox}{\unvbox\@outputbox}{}{%
	\errmessage{\noexpand\@combinedblfloats could not be patched}%
}%
\makeatother

\title[Microlensing of non-radially pulsating stars]{Non-radially pulsating stars as microlensing sources}
\author[]{authors}
\author[Sajadian et al.]{Sedighe Sajadian$~^{1}$\thanks{E-mail: s.sajadian@iut.ac.ir}, Richard Ignace$~^{2}$\thanks{E-mail: ignace@etsu.edu}
	                            \footnotemark[1]\\
	$^{1}$Department~of~Physics,~Isfahan~University~of~Technology,~Isfahan~84156-83111,~Iran\\
	$^{2}$ Department of Physics \& Astronomy, East Tennessee State 
	University, Johnson City, TN 37614, USA}
	                            
\date{Accepted XXX. Received YYY; in original form ZZZ}
\pubyear{2020}

\begin{document}
\label{firstpage}
\pagerange{\pageref{firstpage}--\pageref{lastpage}}
\maketitle
\begin{abstract}
We study the microlensing of Non-Radially Pulsating (NRP) stars. Pulsations are formulated for stellar radius and temperature using
spherical harmonic functions with different values of $l,m$. The characteristics of the microlensing light curves from NRP stars are
investigated in relation to different pulsation modes.  For the microlensing of NRP stars, the light curve is not a simple
multiplication of the magnification curve and the intrinsic luminosity curve of the source star, unless the effect of finite
source size can be ignored. Three main conclusions can be drawn from the simulated light curves. First, for modes with $m\neq0$
and when the viewing inclination is more nearly pole-on, the stellar luminosity towards the observer changes little with pulsation phase.
In this case, high-magnification microlensing events are chromatic and can reveal the variability of these source stars. Second, some
combinations of pulsation modes produce nearly degenerate luminosity curves (e.g., $(l,m)=(3,0), (5,0)$). The resulting microlensing light curves are also degenerate, unless the lens crosses the projected source.  Finally, for modes involving $m=1$, the stellar brightness centre does not coincide with the coordinate centre, and the projected source brightness centre moves in the sky with pulsation phase. As a result of this time-dependent displacement in the brightness centroid, the time of the magnification peak coincides with the closest approach of the lens to the brightness centre as opposed to the source coordinate centre. Binary microlensing of NRP stars and in caustic-crossing features are chromatic.

\end{abstract}

\begin{keywords}
gravitational lensing: micro, Variable stars: non-radial pulsating stars
\end{keywords}

\section{Introduction}

Gravitational microlensing refers to the bending of the light
beam of a background source star due its nearby passage through the
gravitational potential of a foreground body of mass. In this
phenomenon, owing to making two distorted images from the source
star, the source brightness is temporarily magnified \citep{Einstein1936, pac86}.
Gravitational microlensing is not only an astrophysical event, but
also a diagnostic method with many applications, such as detecting
extrasolar planets \citep{Mao1991planet, Gaudi2012}, measuring the
mass distributions in our Galaxy and examining Galactic structures
\citep{moniez, Dominik2006}, and testing candidates for dark matter
such as searching for Massive Compact Halo Objects (MACHOs)
\citep{alcock2000, tisserand2007}.

In addition, microlensing acts as a magnifier that can serve to
enlarge small perturbations in the atmospheres of distant stars,
such as the magnified source stars located in the Galactic bulge
\citep{sackett2001, Heyrovski2000, sajadian2016}. A few examples of
such perturbations over a stellar surface include star spots, stellar
oblateness, or circumstellar disks
\citep{heyrovski1997,han2000,gaudi2004,Rattenbury2005,zheng2005,Rattenbury2009,sajadian2015,Hendry2002}.
This paper seeks to explore such effects in terms of the intrinsic
pulsational variability that many stars experience.

Variable stars are divided into several categories, such as extrinsic
or intrinsic, regular or irregular. These classifications are based
on the variability criteria, periods, the amplitude of variations,
spectral type, and/or luminosity class. Properties of known variable
stars are gathered in the \textit{General Catalogue of Variable
stars} (GCVS) \citep[][]{samus1997,samus2017}
\footnote{\url{http://www.sai.msu.su/gcvs/gcvs/}}. The Optical
Gravitational Lensing Experiment (OGLE) team has identified large
ensembles of different kinds of variable stars both in and out of
our Galaxy \citep{Igor2010, Igor2013, OGLE2013, Igor2014}. Extrinsic
variables are mostly eclipsing binary or planetary systems
\citep{derekas2007, Graczyk2011}.

Among intrinsic variable stars, variation of stellar brightness is
driven by physical phenomena taking place within the star itself,
such as strong magnetism, stellar activity, precession of the
rotational axis in ellipsoidal stars, eruption, etc.
\citep[see, e.g.,][]{Kahn1969,Cox1974,book2007, book2015}.  Generally,
intrinsic variable stars are divided into four subclasses: rotational
variables, eruptive variables, explosive variables and finally
pulsating variables. The latter is further divided into radial and
non-radial pulsating stars. However, pulsating stars can be further
subclassified according to their mass, evolutionary state, etc. The
driver for stellar pulsation is the expansion of the stellar
surface owing to blocking of internal energy flow and then restoring
to its initial form owing to pressure ($p-$mode) or gravity ($g-$mode),
with the cycle repeating in a periodic fashion. Oscillation in
pulsating stars can happen with one fundamental frequency or with
overtones involving several frequencies. For these stars, there is
a relation between their period and mean mass density, so that the
denser stars have shorter pulsational periods \citep[][]{Ritter1881}.

Typically, surveys that monitor a star field for microlensing events
will remove the variable stars to avoid confusing their variations
with the effects of gravitational lensing \citep{Assef2006}.
However, microlensing can magnify small perturbations in stellar
atmospheres.  Consequently,
measurement of the light curves from pulsating variable
stars that are undergoing microlensing can help discern the
variability parameters (e.g., its period, amplitude, etc).  When
combined with asteroseismology, information can be obtained about
the stellar interiors \citep[one example is][]{Varmicro}. In
\citet[][~hereafter Paper~I]{PaperI}, we developed a formalism for
microlensing of radially pulsating stars. We classified
possible perturbations due to stellar pulsation in single and binary
microlensing light curves and discussed the plausibility of determining
the variability parameters in microlensing observations. Here, we
expand that initial work to include non-radial pulsating stars as
microlensing source stars.

The outline of the paper is as follows.  In section \ref{model},
we introduce the formalism  for describing NRP stars used in this
paper. Also we discuss the pulsation of the stellar luminosity
and its shape in different modes.  In section \ref{micro},  we
revise the  properties of microlensing light curves from NRPs in
different pulsation modes. In section \ref{binary} we consider the
binary lensing of NRP stars and investigate the characteristics of
caustic-crossing binary light curves from these stars. 
In the last section we summarize the conclusions.

\begin{figure*}
\centering
\includegraphics[width=0.99\textwidth]{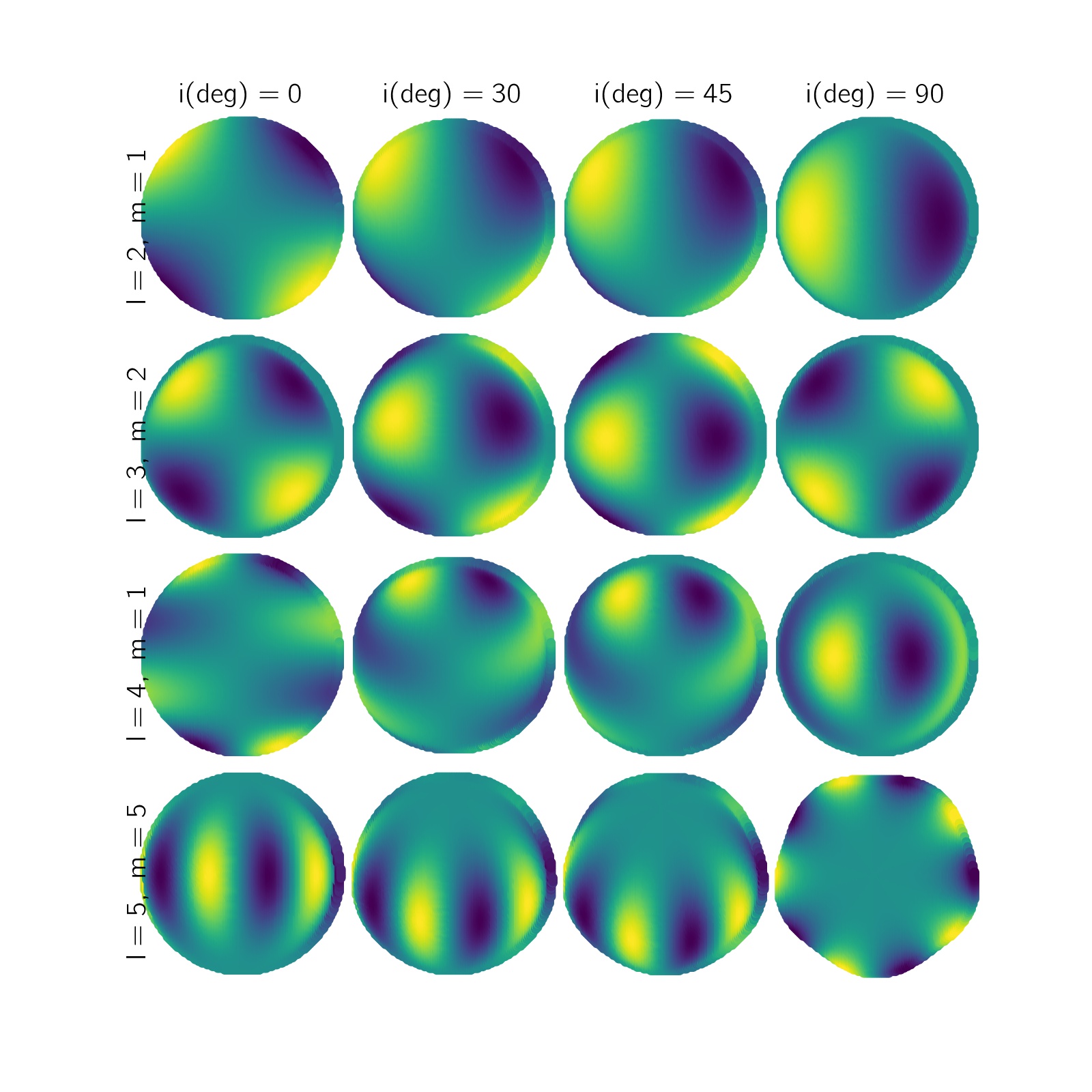}
\caption{Examples of source stars displaying NRPs with different inclination
angles ($\rm{i(deg)}$) and harmonic modes ($\rm{l},\rm{m}$). The
coloured features indicates the variable stellar surface temperatures.
The parameters are $\delta_{\rm{R}}= 0.07~R_{\star}$,
$\delta_{\rm{T}}=200~$K.}\label{star}
\end{figure*}
\section{Modeling non-radial pulsations}\label{model}

The theory of non-radially pulsating stars was first studied by
\citet{Thompson1863}. The theory continued to advance with a focus
on modeling the radial variations from the stellar centre using
spherical harmonic functions \citep[see, e.g., ][]{cowling1941,
Dziem1977}. We adopt these oscillatory functions for modeling the
temperature and radial variations arising from NRPs
with latitude and longitude about the star.  Modification to the
stellar radius, $R$, in a given direction of co-latitude ($\theta_{\star},$
and azimuth $\phi_{\star}$) over the stellar surface, with time
$\rm{t}$, is represented as \citep{Bonanno2004}:

\begin{eqnarray}\label{deltar}
\Delta R(\theta_{\star}, \phi_{\star})= \bar{R}~\delta_{R}~c_{\rm l,m}
	~P_{\rm l,m}(\cos \theta_{\star})~\cos [ \omega (t-t_{\rm p})
	+m\phi_{\star}],	
\end{eqnarray}	

\noindent where $P_{\rm l,m}(\cos \theta_{\star})$ is the associated
Legendre polynomial, 

\begin{equation}
c_{\rm l,m}^{2}=\frac{2l+1}{4
\pi}\frac{(l-m)!}{(l+m)!}
\end{equation}

\noindent  is the normalization factor of the
spherical harmonics, $\delta_{\rm{R}}$ is the amplitude of pulsation
over the source surface which is normalized to $\bar{R}$, the mean
value of the stellar radius over one pulsational cycle.
Note that $\omega=2 \pi/ P$ is the angular velocity associated with the
pulsation period $P$, and $t_{\rm p}$ is an arbitrary time. We note
that any local change in the stellar radius causes a corresponding
variation in the stellar temperature, $\Delta T$, which we formulate
as follows:

\begin{eqnarray}
\Delta T= \delta_{T}~c_{l,m}~P_{l, m}(\cos \theta_{\star}) \cos [\omega 
	(t-t_{p}) + m \phi_{\star}+ \phi_{0}],
\end{eqnarray}

\noindent where, $\delta_{\rm{T}}$ is the amplitude of variation
in the surface temperature due to the stellar pulsation, and $\phi_{0}$
is the phase difference between the variation in the stellar radius
and its surface temperature. We fix $\phi_{0}=\pi/2$
\citep{Carrollbook}.

\begin{figure*}
\centering
\subfigure[]{\includegraphics[angle=0,width=0.49\textwidth,clip=0]{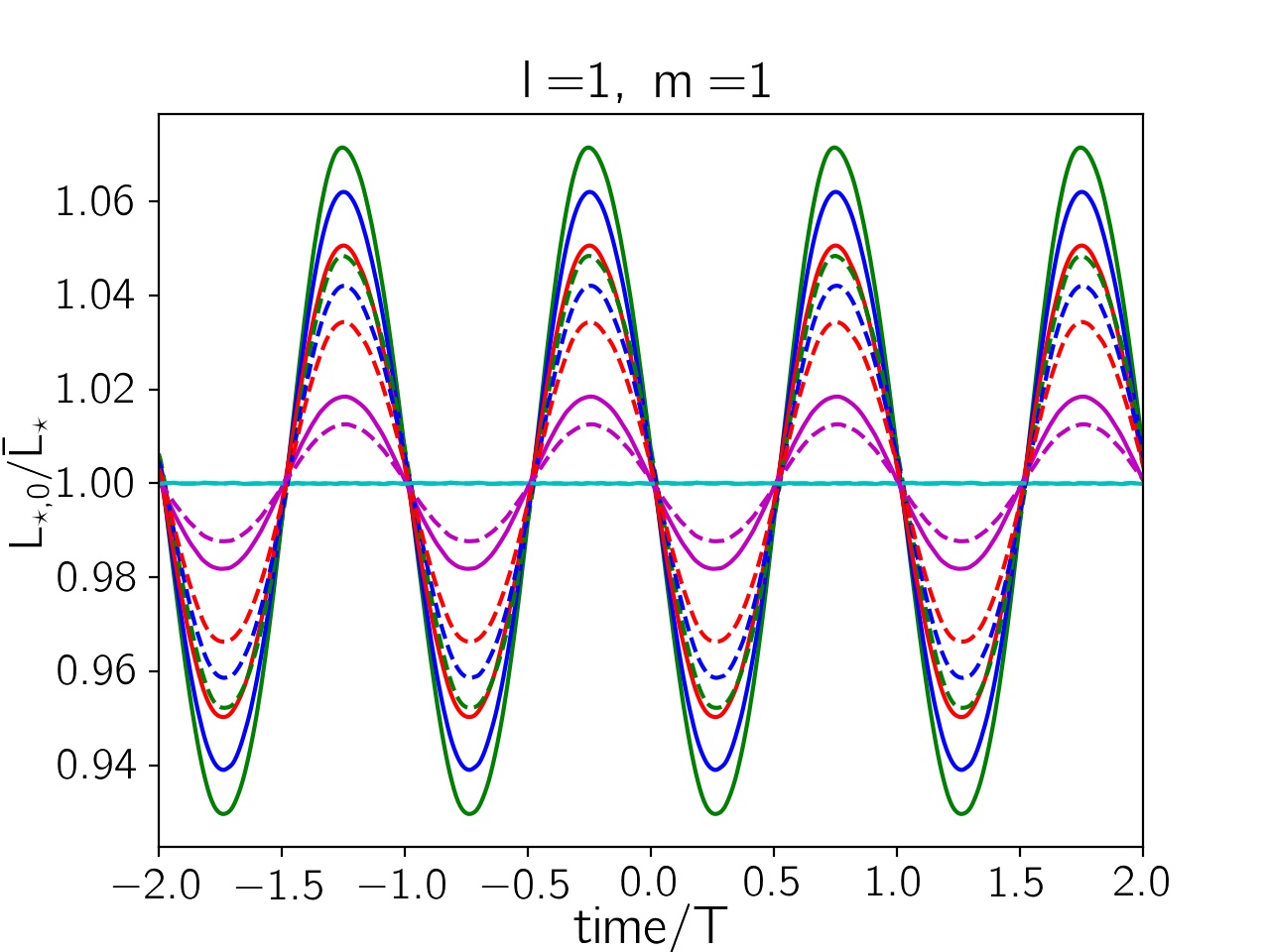}\label{la}}
\subfigure[]{\includegraphics[angle=0,width=0.49\textwidth,clip=0]{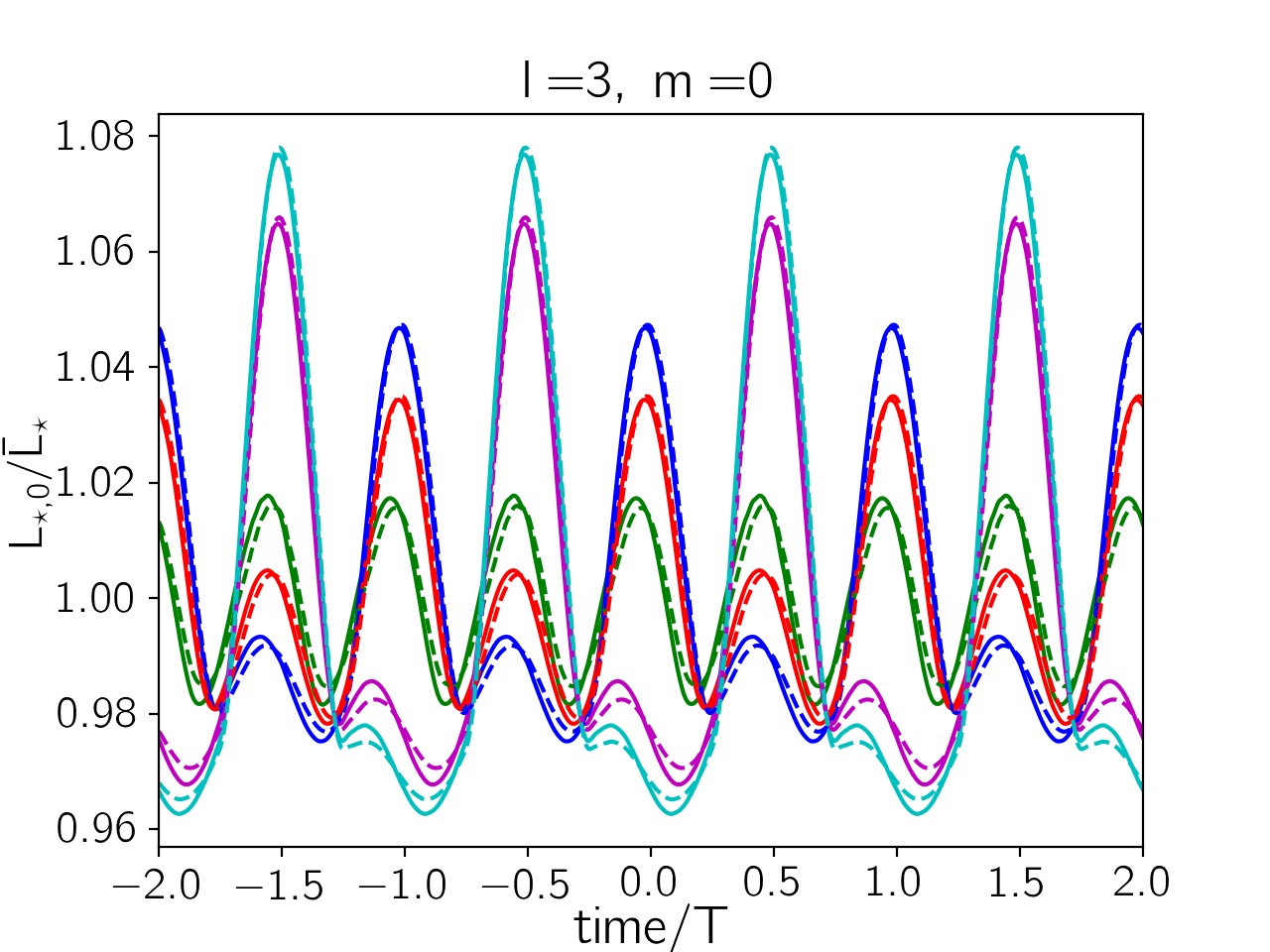}\label{lb}}
\subfigure[]{\includegraphics[angle=0,width=0.49\textwidth,clip=0]{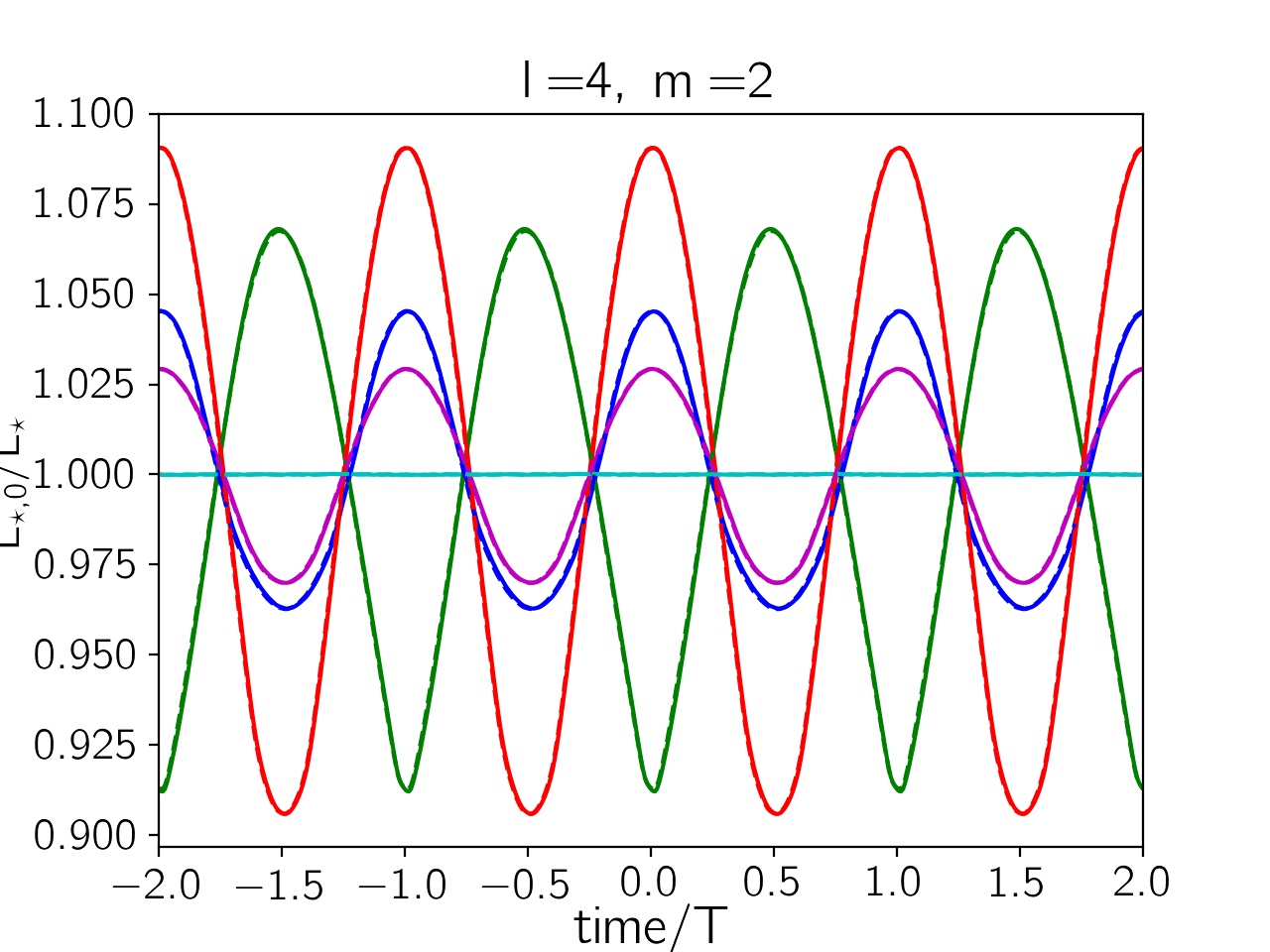}\label{lc}}
\subfigure[]{\includegraphics[angle=0,width=0.49\textwidth,clip=0]{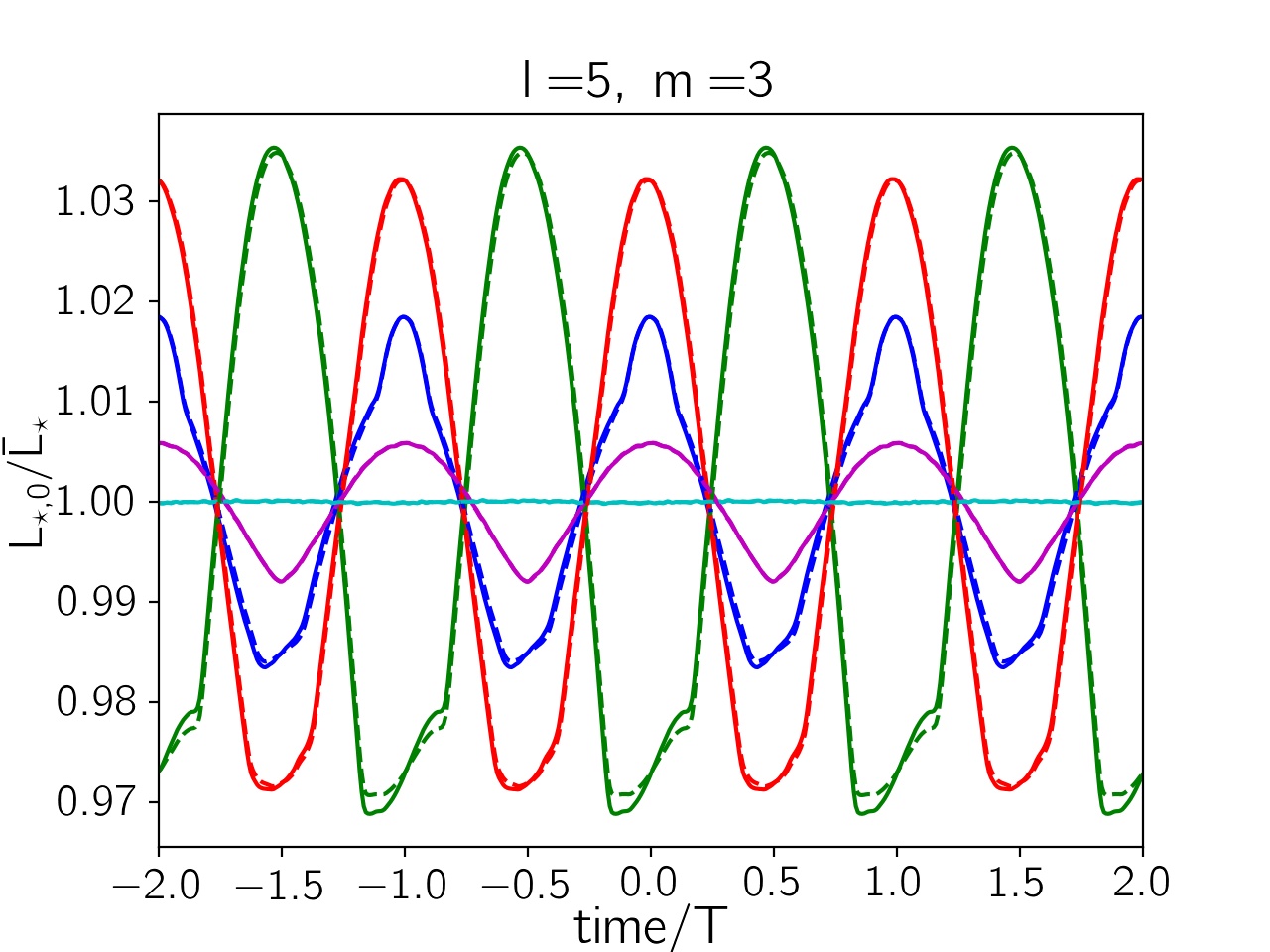}\label{ld}}
\caption{Four example lightcurves for non-radially pulsating stars
with different pulsational modes, $(l,m)=(1,1), (3,0),  (4,2),
(5,3)$.  Lightcurves are shown as passband stellar luminosities
normalized to cycle averages $\bar{L}_{\star}$. Modes are identified
at the top of each panel. Curves with inclination angles
$i=0,~30,~45,~75,~90~$deg are shown as green, blue, red, and magenta,
cyan colours, respectively. The luminosities in $V$ and $I-$bands
are plotted as solid and dashed curves, respectively. For these
plots, the parameters of the source stars are  $\delta_{\rm T}=450$T,
$\delta_{\rm R}=0.35\bar{R}$, and $\phi_{0}= \pi/2$.}\label{lumino}
\end{figure*}

For the observer, the variation in brightness will depend on viewing
inclination in relation to the projected pattern of NRPs.
We introduce two coordinate systems. The first
one is a source system for the star, with cartesian coordinates
($x_{\star},~y_{\star},~z_{\star}$), whose centre is the source
centre. Each point over the source surface in this coordinate is
specified as:

\begin{eqnarray}
x_{\star}&=& R \sin \theta_{\star} \cos \phi_{\star},\nonumber\\
y_{\star}&=& R \sin \theta_{\star} \sin \phi_{\star},\nonumber\\
z_{\star}&=& R \cos \theta_{\star},
\end{eqnarray}

\noindent where $R = \bar{R}+ \Delta R(\theta_{\star}, \phi_{\star},t)$.
The second system is that of the observer, with cartesian coordinates 
($x_{o},~y_{o},~z_{o}$). In this system, the $x_{o}$-axis is towards the
observer and normal to the sky plane. the $y_{o}-z_{o}$ plane indicates
the sky plane, and its centre coincides with the centre of the source
coordinate system. One can convert the first system to the second
by a rotation around $y_{\star}$ with inclination angle
$i$, as follows:

\begin{eqnarray}
x_{o}&=&\hphantom{-}x_{\star} \cos i  +  z_{\star} \sin i,\nonumber\\
y_{o}&=&\hphantom{-}y_{\star}, \nonumber\\
z_{o}&=&-x_{\star} \sin i + z_{\star} \cos i.
\end{eqnarray} 

At any moment only those points $x_{o} \ge 0$ can be seen by the
observer and constitute the projected stellar surface on the sky
plane. In Figure (\ref{star}), we show several examples of non-radially
pulsating stars as projected onto the sky
as would be seen by a distant observer.  From top to bottom, four
different harmonic modes are displayed: $\rm{(l, m)}=(2,~1), (3,~2),
(4,~1), (5,~5)$.  The columns moving from left to right are four
different inclination angles:  $i=0,~30,~45, ~90~$deg.  The coloured
features show the variation of the surface temperature. For these
stellar pulsational modes, 
we chose $\delta_{R}=0.07~\bar{R}$, $\delta_{T}=200~$K,
$\phi_{0}=\pi/2~$, $t=t_{\rm{p}}$ and $\bar{R}=1.8~R_{\odot}$. In
these plots the $y_{o}$-axis is towards the right and $z_{o}$ is 
up. Rotation by the inclination angle $i$ is done around
the $y_{\star}=y_{o}$-axis.

NRPs lead to 3 variations as perceived by an observer.  The first
is the emergent stellar luminosity in any given direction.  Second
is the projected area of the source star owing to the complex variations
of radius with latitude and longitude.  And third is the shape of
the star.  This is different from the projected area, as the shape
represents alterations to the boundary of the star in the $y_{o}-z_{o}$
plane (e.g., the shape could change even if the area were constant;
for NRPs, both are changing, whereas for radially pulsating stars
only area changes with time). These three variations are studied
in the following subsections.

\subsection{Luminosity variations from NRPs}\label{lumin}

The luminosity of a star undergoing NRPs as a function of time is
found from integrating the stellar intensity over the projected
stellar surface visible to the observer. Similar to the formalism
which was introduced in Paper~I, we assume that each element of the
stellar surface radiates as a blackbody with the intrinsic surface
intensity, $B_{\lambda}[T_\ast(\theta_\star, \phi_\star)]$. The integration
becomes:

\begin{eqnarray}\label{lumi}
L_{\star,0}(t) & = & \int_{0}^{\infty}d\lambda~K(\lambda-\lambda_{0}) 
	\times \nonumber \\
	& & \int~dS_{o}~B_{\lambda}[\bar{T_\ast}+\Delta T_\ast(\theta_{\star},
	\phi_{\star},t)],
\end{eqnarray}

\noindent where the index $0$ signifies that no lensing effect is
involved yet, consequently $dS_{o}=dz_{o}~dy_{o}$ is the area of
each element on the stellar surface projected onto the sky plane;
$K(\lambda-\lambda_{0})$ is the throughput function for the passband
filter under consideration; and $\bar{T_\ast}$ is the average value of
the stellar surface temperature. The integration is complicated by
the fact that the star is misshapen and not circular in projection.
For this task, we apply the method of 'inverse ray shooting'
\citep{ray2, ray1}.

For four non-radially pulsating stars with different inclination
angles, $(l,m)=(1,1), (3,0),  (4,2), (5,3)$, their normalized
luminosity curves, $L_{\star,0}(t)/\bar{L}_{\star}$, are plotted
in Figure~(\ref{lumino}) in standard filters $V$ (solid curves) and
$I$ (dashed curves). Here, $\bar{L}_{\star}$ is the average value
of the stellar luminosity over one pulsational period, $T$:

\begin{equation}
\bar{ L}_{\star} =\frac{1}{T} \, \int_{t}\,  L_{\star, 0} \, dt.
\end{equation}

\noindent Lightcurves for inclination angles $0,~30,~45,~75,~90$deg
are displayed with green, blue, red, magenta and cyan colours,
respectively.  The variations of the stellar radius and its surface
temperature drives changes in the stellar luminosity with time.

\begin{figure}
\centering
\includegraphics[width=0.49\textwidth]{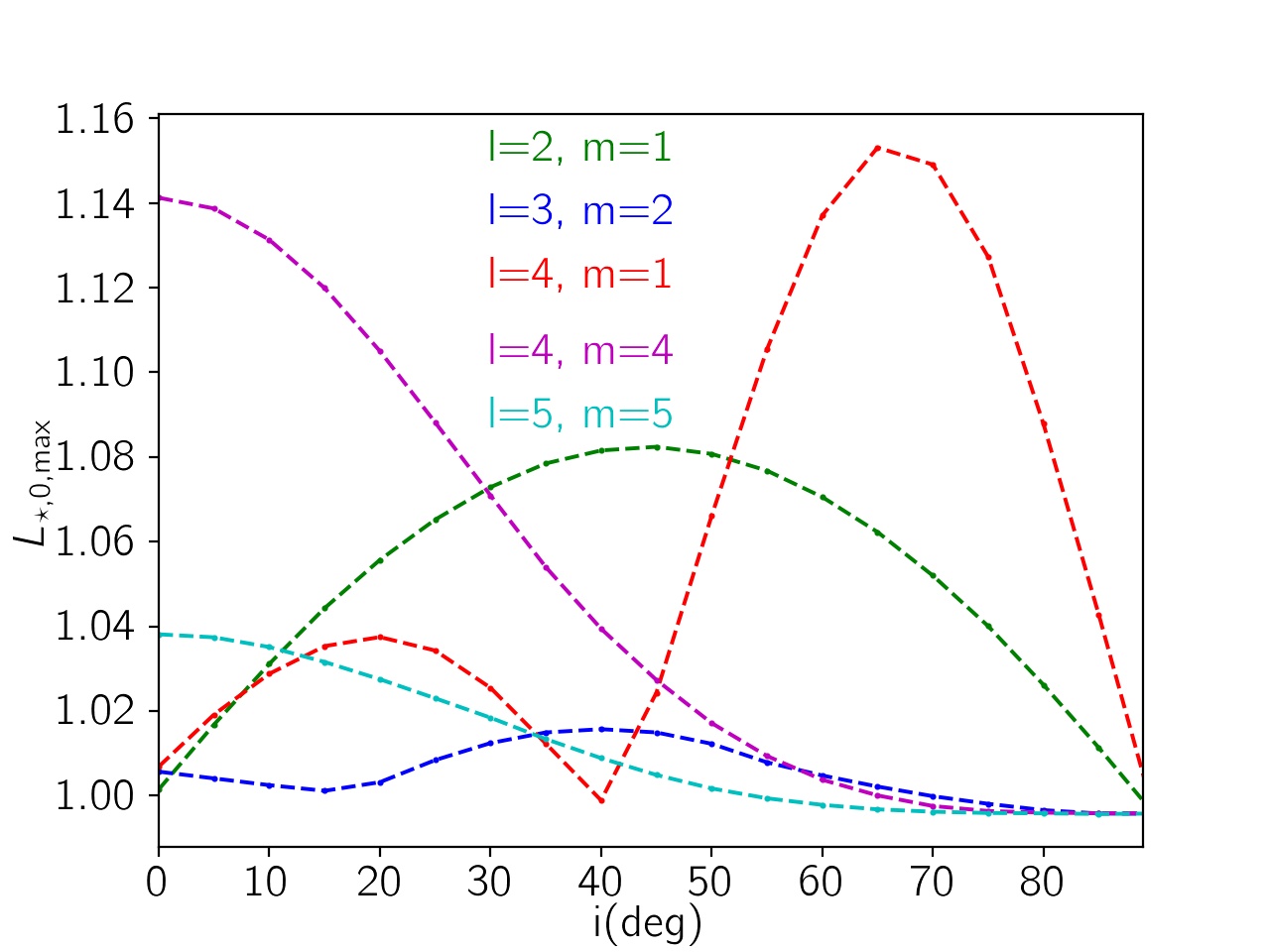}
\includegraphics[width=0.49\textwidth]{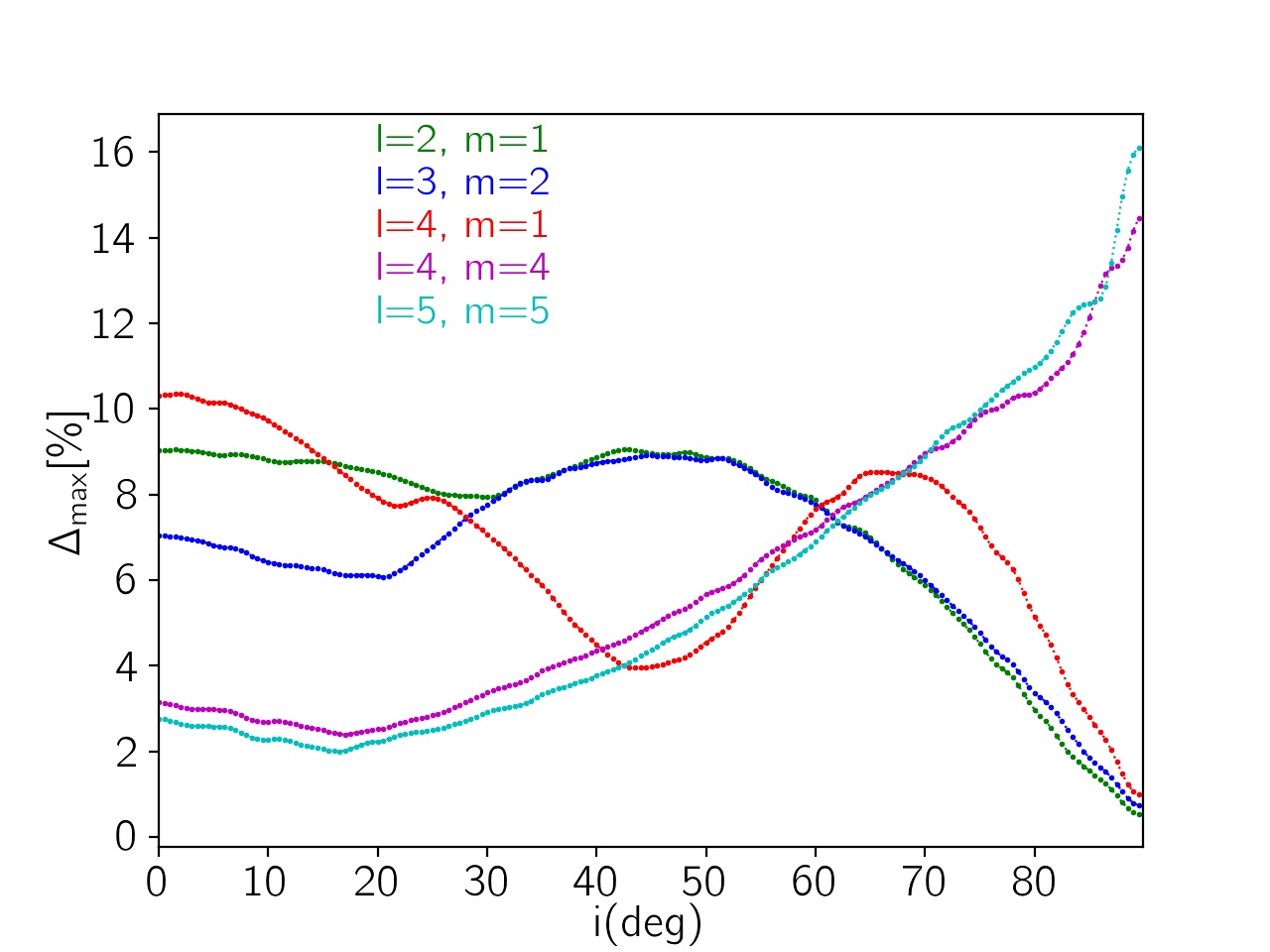}
\caption{The amplitude of the stellar intrinsic lightcurve (top)
and the reshape factor (bottom; see eq.~[\ref{reshape}]) for harmonic
modes with $(l,m)= (2,1)$, $(3,2)$, $(4,1)$,  $(4,4)$,  $(5,5)$
versus the inclination angle, which are plotted with green, blue,
red, magenta and cyan colours, respectively.}\label{cparam}
\end{figure}

Key points regarding these curves are listed as follows:

\begin{enumerate}[label= \arabic*., leftmargin=0.1cm]  
	
\item All curves have periodic behavior, but the variations are not
simple sinusoidal forms, except for modes $(l,m)=(0,0), (1,1),
(2,1), (2,2)$. Hence, these modes are degenerate and discerning their pulsation modes, when the stars are distant, essentially point-like, is impossible. Figure \ref{la} presents the lightcurves for a NRP
star with $(l,m)=(1,1)$ that shows the lightcurve to have
a simple sinusoidal function.\\

\item For $i=90~$deg, the observer sees the $x_{\star}-y_{\star}$
plane of the star. In this case the projected surface of the source
star as seen by the observer will appear to rotate around $x_{o}$ 
(for $m \neq 0$), with $m$ rotations per pulsational period.  The
rotation of the source surface does not alter the luminosity as
seen by the observer; as for example in Figures \ref{la}, \ref{lc}, \ref{ld}.\\

\item When $m=0$ and the stellar pole is towards the observer, 
the net stellar surface brightness decreases by the
factor $\cos[\omega (t-t_{\rm p}) + \phi_{0}]$ and causes a periodic
pulsation with the largest amplitude (with respect to other
value of the inclination angle).  An example for the mode $(l,m)=(3,0)$
is shown in Figure \ref{lb}. We note that generally in modes
with $m=0$, the luminosity curves for any values of the inclination
angle vary in a periodic manner with nonzero amplitude. \\

\item The variations of the stellar luminosity in different filters
for NRP stars are not as high as for stars that are radial
pulsators. For NRPs, sectors with increased and decreased temperatures
lead to a surface-averaged value that has less variation in comparison
to a star undergoing radial pulsations. However, for the modes
$(l,m)=(0,0), (1,0), (1,1)$, the stellar luminosity in different
filters can have considerable differences, as in the example
of Figure \ref{la}.\\

\item The amplitude of the lightcurves in different pulsation modes
generally depends on the inclination angle. The maximum amplitude
of variation is achieved at viewing inclinations associated with
the stellar latitude at which the Legendre polynomial displays a
maximum.  If that latitude is $\theta_\star$, then the variation
is maximized for inclination $i=\pi/2 - \theta_\star$. Because when
the stellar surface is projected on the sky plane, the points over
the equator have $\theta_{\star}=\pi/2-i$. These points have the
largest impact on the stellar luminosity.  As an example, the
Legandre function for the $(3,1)$ mode is proportional to $\sin
\theta_{\star}[5\cos^{2}(\theta_{\star}) -1]$ which is maximized
for $\theta_{\star}=31~$deg.  Hence, the maximum amplitude of the
luminosity happens for $i=59~$deg. \\

\item We study the luminosity curves for all pulsation modes with
$l<6$.  In order to compare the amplitudes of the luminosity curves,
we fix the pulsation parameters at $\delta_{\rm T}=450$K and
$\delta_{\rm R}=0.35\bar{R}$ when $\phi_{0}=\pi/2$.  In the top panel of
Figure \ref{cparam}, we show the amplitude of the luminosity curve
of NRPs versus the inclination angle for the following pulsation modes,
$(l,m)= (2,1)$, $(3,2)$, $(4,1)$,  $(4,4)$, $(5,5)$ with green,
blue, red, magenta and cyan colours, respectively. Accordingly, the
maximum amplitude for the luminosity curve was for the mode
$(l,m)=(0,0)$ for all inclination angles which is exactly a radial
pulsating star. The second and third maximum amplitude of the
luminosity curves is for the mode $(l,m)=(4,0), (2,0)$ at the
inclination angle $i=90~$deg, (pole-on view).\\

\item For a fixed value of $l$, the amplitude of the
luminosity curves decreases by increasing the $m$ value from zero to
$l$.  This means that for all modes with fixed $l$,  the mode with
$m=0$ produces the largest amplitude for the intrinsic luminosity curve.

\end{enumerate}

\subsection{Shape variations from NRPs}\label{reshap}

For a star undergoing NRPs, the shape of the star as seen by the
observer is not always circular. The degree of asymmetry in its
shape depends on both the inclination angle and the pulsation mode
($l,~m$), and the amount of distortion
scales with the factor $\delta_{\rm{R}}$.
In order to study the amount of asymmetry in the shape of 
stars undergoing NRPs, we define a \textit{reshape factor} as:

\begin{eqnarray}\label{reshape}
\Delta=\frac{\delta R}{\bar{R}}=\frac{ \stretchleftright[1000]{\langle}
	{\begin{array}{c}|R_{\star}(\theta_{\star},\phi_{\star}| x_{o}=0)-\bar{R}| \end{array}}
	{\rangle} }{\bar{R}},
\end{eqnarray}

\noindent which is the average of the positive-definite relative
difference of the stellar radius of each point at the stellar edge
(for which $x_{o}$ is zero) from the average source radius.  For
these points we have $\tan(i)+\tan(\theta_{\star}) \cos(\phi_{\star})=0$.
This parameter is a measure for how much the stellar shape differs
from a circle of average radius $\bar{R}$. For harmonic
functions with $(l,m)= (2,1)$,  $(3,2)$, $(4,1)$, $(4,4)$  and
$(5,5)$, we evaluate $\Delta_{\rm{max}}$
numerically as a percent with results shown in the bottom panel of
Figure~(\ref{cparam}), using $\delta_{\rm R}=0.35$.
Comparing the magnitudes of the reshape factor for different
modes is meaningful here, because $\delta_{\rm R}$ is fixed for all
the modes. We note that the reshape factor is generally time-dependent
and varies periodically.  The following are several key points
based on Figure~(\ref{cparam}).

\begin{enumerate}[label= \arabic*., leftmargin=0.1cm]

\item Generally, when the polar axis of the source star is directed
towards the observer ($i=90$), for pulsating modes involving Legendre
polynomials that are proportional to $\cos^{n} \theta_{\star}$ (for
any value of $n$), there is no variation in the projected area of
the source star. This arises because
the observer views the $x_{\star}-y_{\star}$
plane, and the edge points have $\theta_{\star}=90~$deg. The relevant modes
are $(l,m)=(1,0)$, $(2,1)$, $(3,0)$, $(3,2)$, $(4,1)$, $(4,3)$,
$(5,0)$, $(5,2)$ and $(5,4)$.  \\

\item When the inclination angle is $i=90~$deg, the modes with $l=m$
produce the largest reshape factors. For $\delta_{\rm R}=0.35$, the
reshape for these modes is $12\%-14\%$.  For these modes
$P_{l,l}(\cos \theta_{\star}) \propto \sin^{l} \theta_{\star}$,
giving an amplitude of variation in the stellar radius that is
maximum (see the last row of Fig.~\ref{star}). By contrast, 
observing at $i=0$~deg yields reshape factors with very small values
for these same modes.\\

\item The local maxima of the  amplitudes of the reshape function
for each mode correspond to where the points over the stellar pole
as seen by the observer (i.e., $\theta_{\star}= i, \pi-i$) have
the maximum domain of the variation.  This condition corresponds
to the Legendre function having a maximum.  For instance,
$P^{2}_{1}=\sin\theta_{\star} \cos \theta_{\star}$ is maximized at
$\theta_{\star}=45~$deg.  Hence its reshape factor is likewise
greatest at $i=45~$deg (green curve in the bottom panel of
Fig.~\ref{cparam}). \\

\item The local minima of the amplitudes of the reshape function
occur when the stellar points over the stellar equator as seen by
the observer (i.e., $\theta_{\star}= \pi/2-i$) have the maximum
domain of the variation.  This condition corresponds to the maximum
value for the Legendre function. For instance,
$P^{3}_{1}=\sin\theta_{\star} ( 5\cos^{2} \theta_{\star}-1)$ is
maximized when $\theta_{\star}=31~$deg.  Hence its reshape factor
is likewise greatest at $i=31~$deg. Correspondingly, its minimum
value will be achieved for $i=59~$deg.\\

\end{enumerate}

\subsection{Areal variations from NRPs} 

NRPs also cause the source area to evolve over time. The enhancement
in the projected stellar area is maximum for modes with $l$ even,
$m=0$, and pole-on views ($i=90^\circ$).  Then the source star has
circular symmetry, with a radius given by $\bar{R}+ \delta_{R}
P^{l}_{0}$).  For these modes and for the pole-on view,  the amplitude
of the stellar luminosity maximize as well. The largest expansion
happens for the mode with $l=m=0$ which is just a radially pulsating
star.  Generally, when the source area is enhanced,  its luminosity increases as well. 

In order to investigate the effects of microlensing of stars with
NRPs, we now include the magnification factor for each element of
source surface while evaluating the stellar luminosity in equation
(\ref{lumi}). The magnified luminosity for a star undergoing NRPs
is given by:

\begin{eqnarray}\label{lstar}
L_{\star}(t) & = & \int_{0}^{\infty}d\lambda~K(\lambda-\lambda_{0})\int~dS_{o}~B_{\lambda}(\bar{T}+\Delta T(\theta_{\star},	\phi_{\star},t)) \nonumber\\
	             & & \times A(u_{\rm y},u_{\rm z}).
\end{eqnarray}

\noindent Here, $A(u_{\rm y},u_{\rm z})$ is the magnification factor of
each element of the source surface. The coordinates $(u_{\rm y},u_{\rm z})$
specify the normalized position of each areal element on the source
surface with respect to the lens location, as given by:

\begin{eqnarray}
u_{\rm y}&=& u_{l,\rm y} - y_{o} /R_{\rm{E},\star} , \nonumber\\
u_{\rm z}&=& u_{l, \rm z} - z_{o}  /R_{\rm{E},\star},
\end{eqnarray}

\noindent where $R_{\rm{E,\star}}=R_{\rm{E}} D_{\rm{s}}/D_{\rm{l}}$
is the Einstein radius as projected onto the source plane. Consequently,
$(u_{l,\rm y}, u_{l, \rm z})$ specify the projected position of the
lens with respect to the projected source centre and are normalized
to the Einstein radius. We define the observed magnification factor
$A_o$ as the magnified stellar luminosity $L_{\star}$ normalized
to the stellar average obtained over a full pulsation cycle:

\begin{eqnarray}
A_{o}=\frac{L_{\star}(t)}{\bar{L}_{\star}}.
\end{eqnarray}

\noindent In this paper, our goal is to study the behavior of the
time-dependent total magnification $A_o(t)$ for different NRP modes
and classify the perturbations induced by stellar pulsation in the
microlensing light curves. \\

\begin{figure*}
	\centering
	\subfigure[]{\includegraphics[angle=0,width=0.49\textwidth,clip=0]{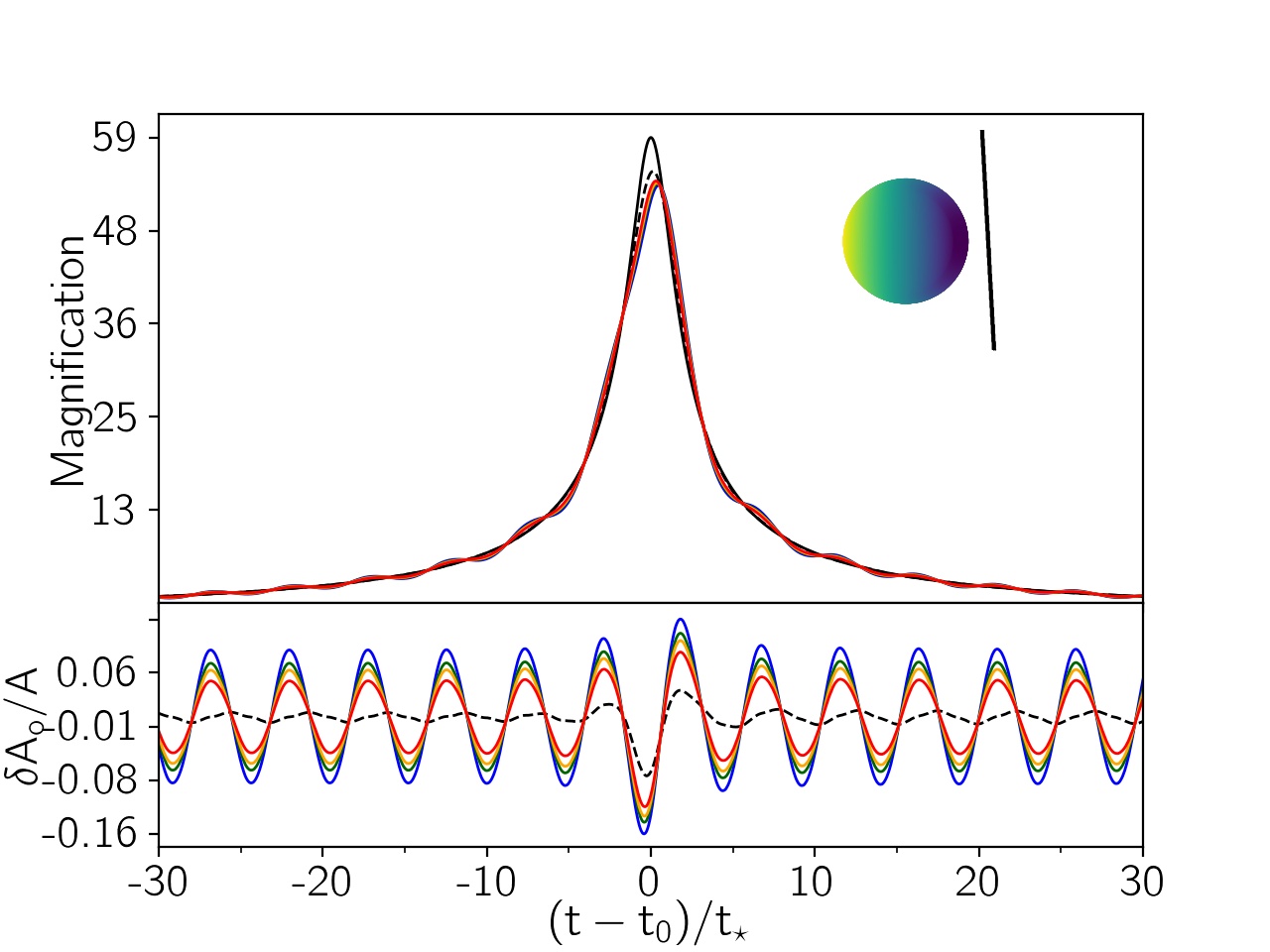}\label{figa}}
	\subfigure[]{\includegraphics[angle=0,width=0.49\textwidth,clip=0]{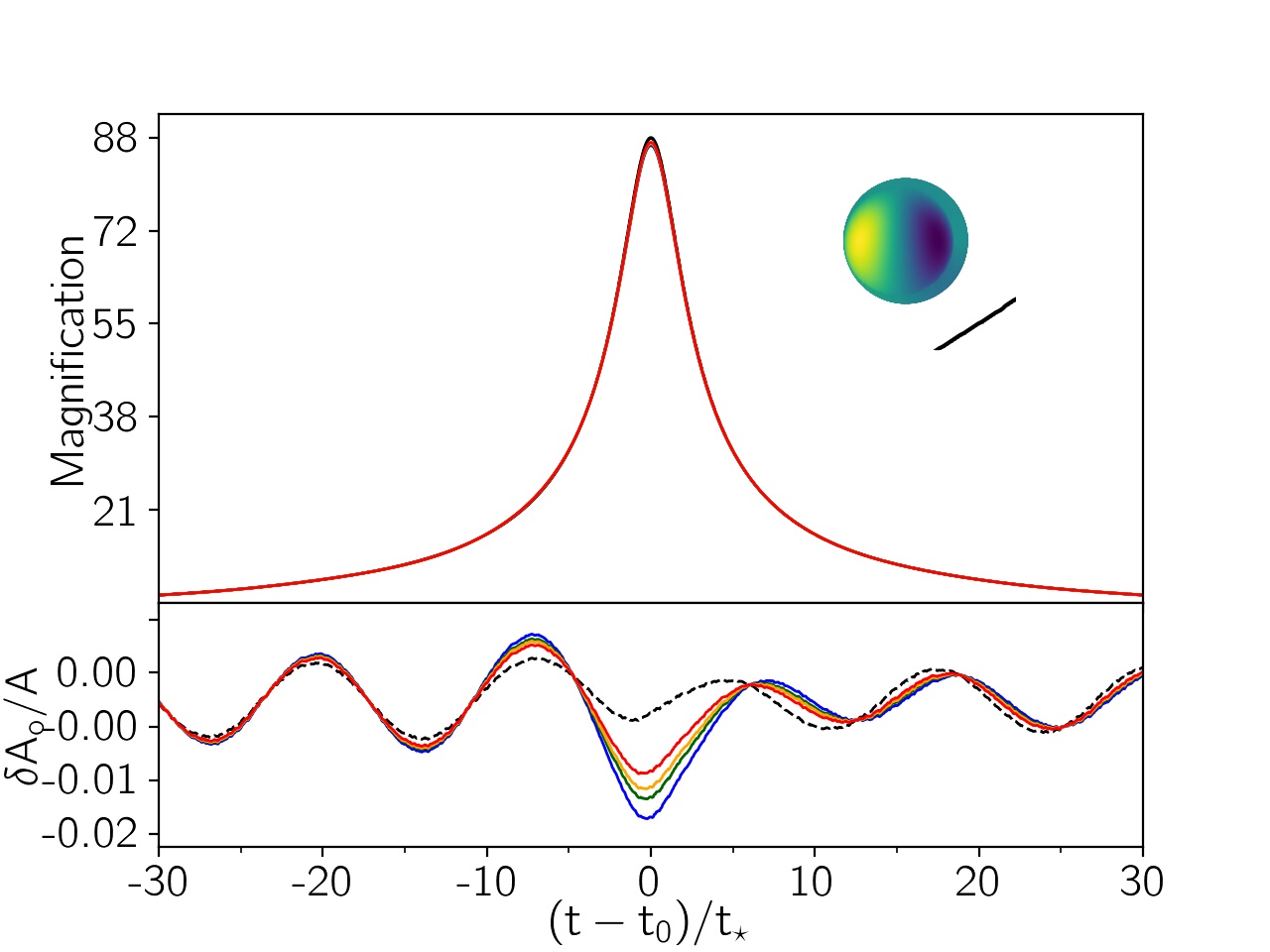}\label{figb}}
	\subfigure[]{\includegraphics[angle=0,width=0.49\textwidth,clip=0]{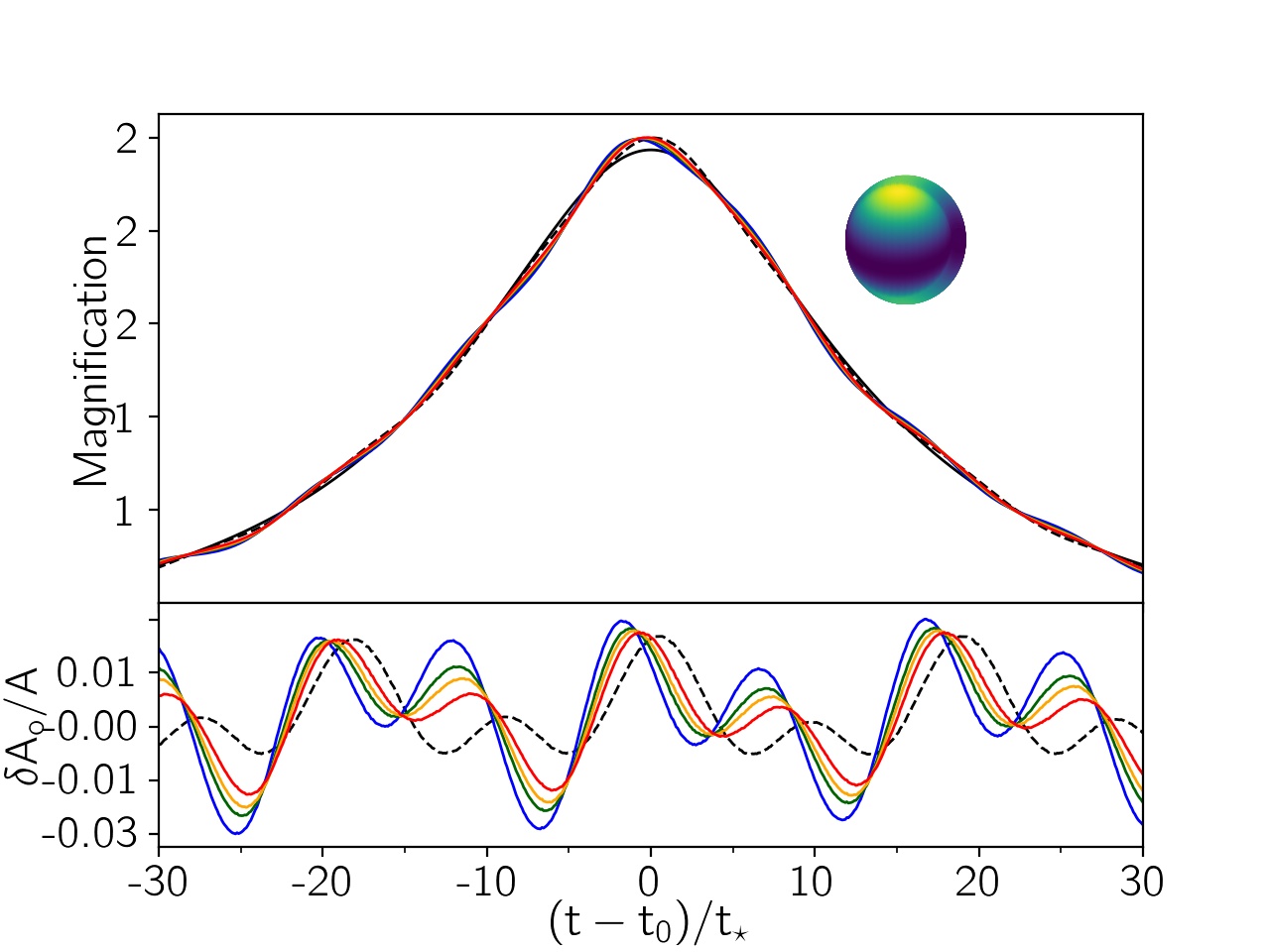}\label{figc}}
	\subfigure[]{\includegraphics[angle=0,width=0.49\textwidth,clip=0]{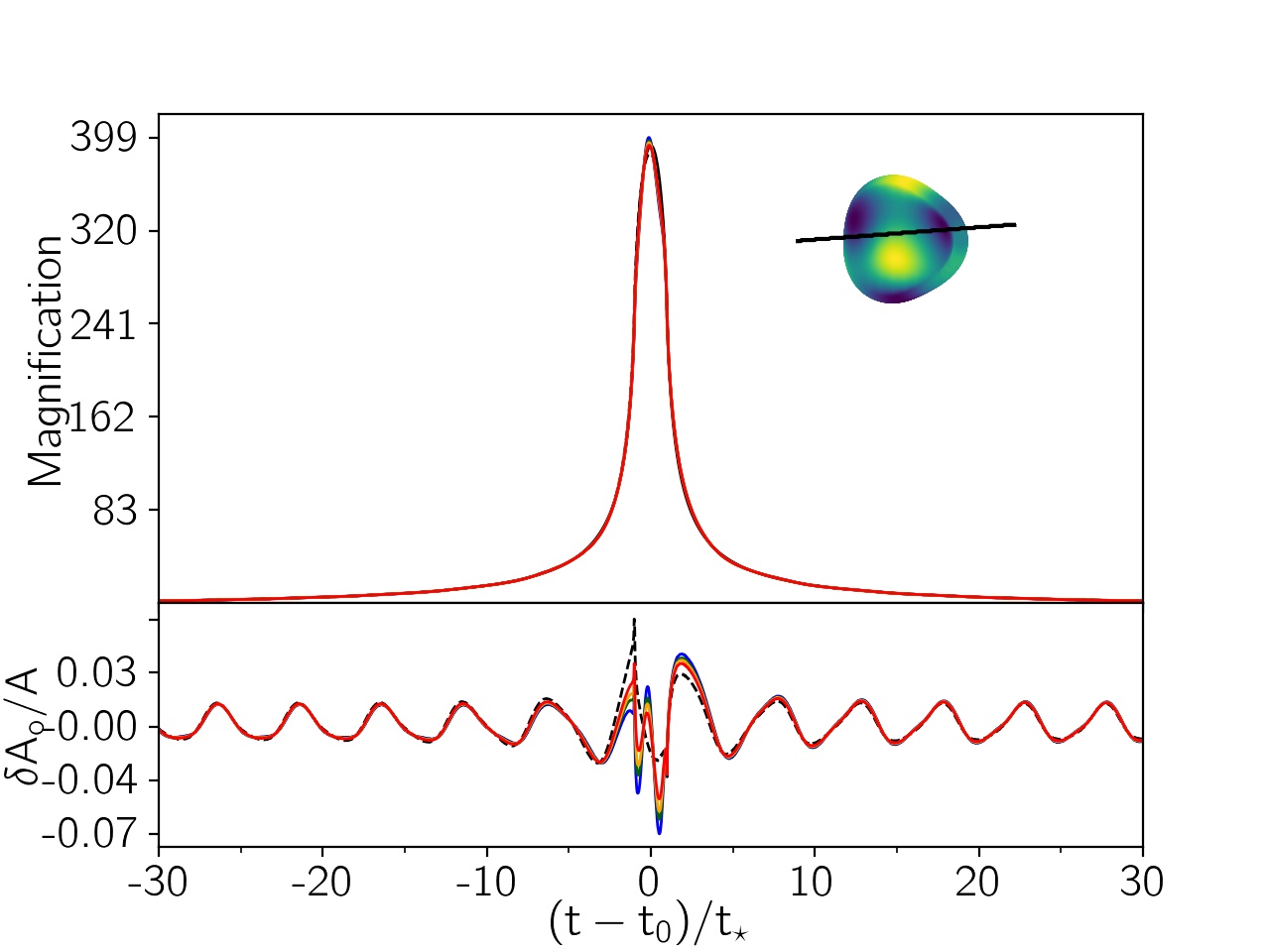}\label{figd}}
	\subfigure[]{\includegraphics[angle=0,width=0.49\textwidth,clip=0]{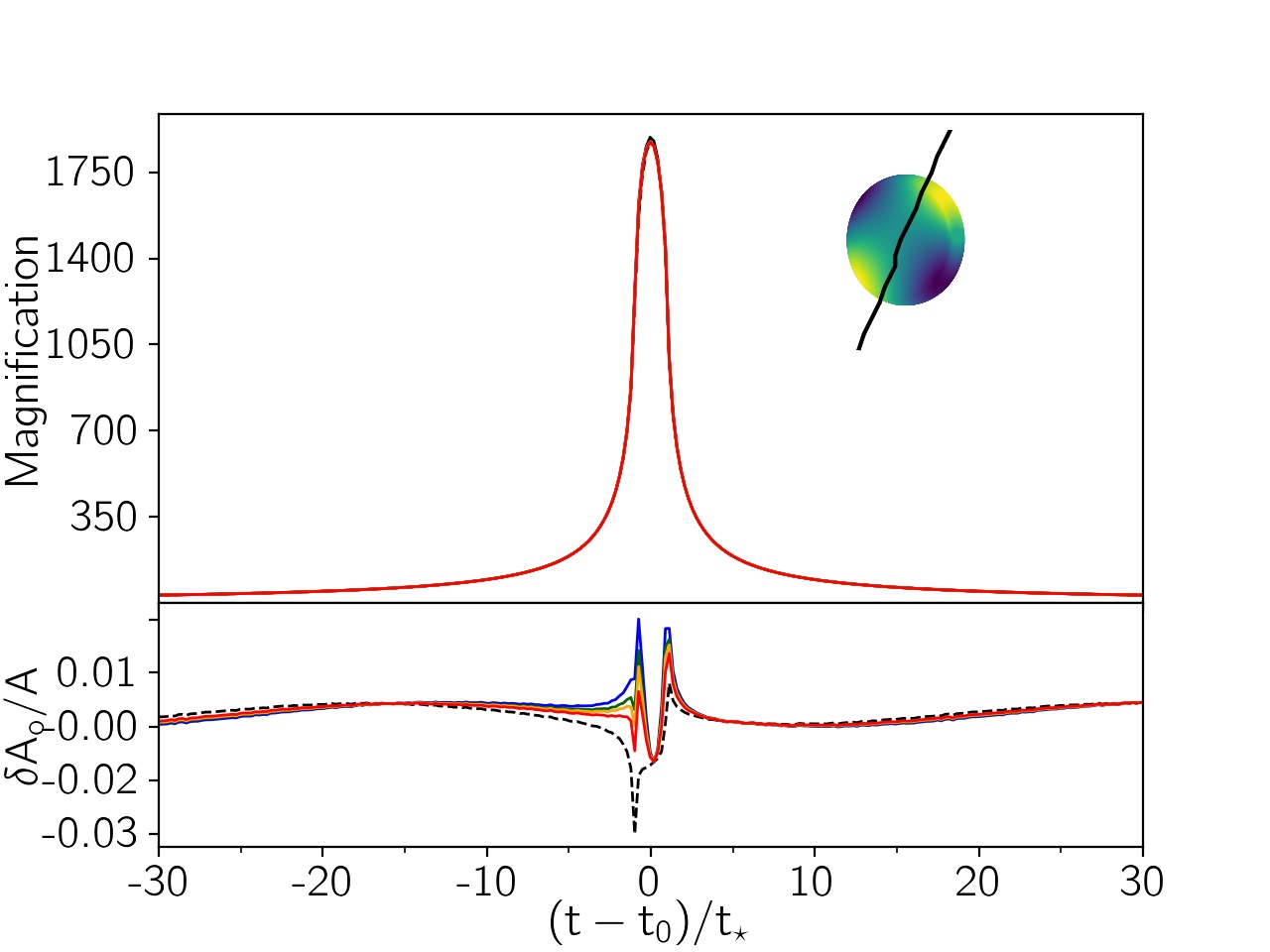}\label{fige}}
	\subfigure[]{\includegraphics[angle=0,width=0.49\textwidth,clip=0]{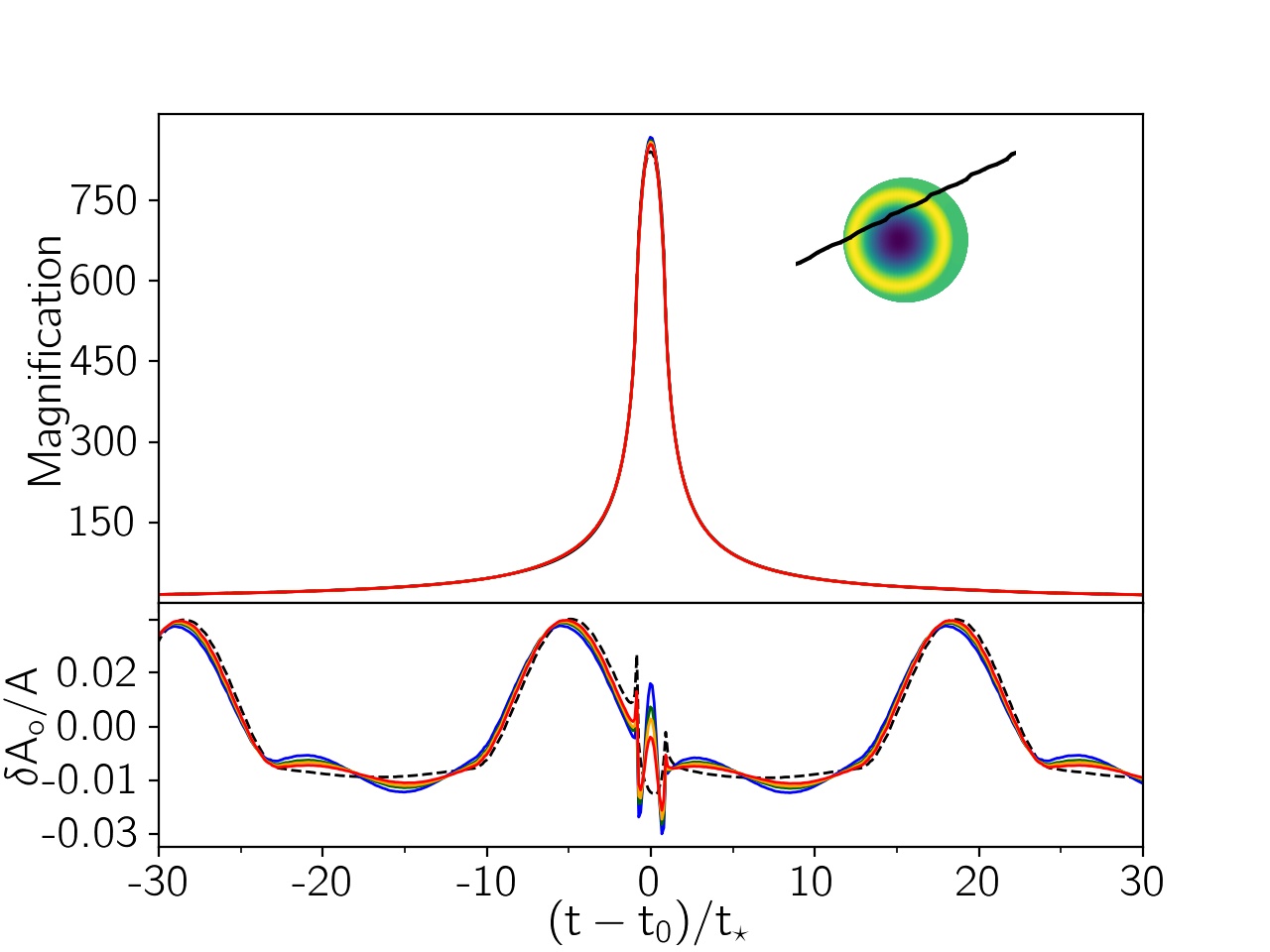}\label{figf}}
\caption{Several examples of microlensing light curves involving
NRPs and a single lens.  Reference microlensing light curves for
no pulsations and using a source star with radius $\bar{R}$ are
shown as solid black curves. Light curves from stars with NRPs, but
suppressing the effect of temperature variations, are shown as
dashed black curves. Light curves involving both variable radius
and temperature for standard passband filters BVRI are plotted as
blue, dark-green, orange and red solid curves.  Lower subpanels plot
a relative difference in relation to the non-pulsating case (solid
black curves).  Insets represent the source
star as seen by an observer at the start time of the light curve,
along with the lens trajectory projected into the source plane.
}\label{light1}
\end{figure*}

\section{Single microlensing of non-radially pulsating stars}\label{micro}

The first microlensing event of a variable source star was discovered
in 1997 towards the Small Magellanic Clouds (SMC), namely
$\rm{MACHO}$-$97$-$\rm{SMC}$-$1$ \citep{alcock1997}. In this event
the variability was used to break the blending-parallax
degeneracy \citep{Assef2006}. In their microlensing survey, the OGLE
collaboration probed the microlensing candidates of variable
source stars and discovered 137 candidates \citep{Lukaz2006}.
Recently, another microlensing candidate from a variable and bright
source star was reported. For this event, the asteroseismic analysis
was done to find the intrinsic variability curve of the source star
from the baseline data \citep{Varmicro}.  These candidates reveal
the importance of investigating the characteristics of microlensing
light curves of variable starts.

For this study we consider the influence of microlensing for stars
undergoing NRPs.  A large ensemble of simulated events have been
explored.  Parameters for the lens and source stars are chosen using
distribution functions described in previous papers 
\citep[c.f., ][]{sajadian2019}. The pulsation parameters are chosen uniformly
in these ranges: the pulsation period  $T\in [0.1,~7]~$days, $\omega
t_{\rm p} \in [0 ,~2\pi]$,  $\delta_{\rm T}\in [250,600]~$K,
$\delta_{\rm R}\in [0.1,0.35]\bar{R}$. The simulation of the
microlensing light curves are done in the time interval $[-40,
40]t_{\star}$, where $t_{\star}=t_{\rm E}\rho_{\star}$ is the
crossing time of the source radius by the lens.

In Figures (\ref{light1}) and (\ref{light2}), several examples of
microlensing light curves from NRPs with different modes are shown.
In these plots, the microlensing light curves without pulsational
effect from a source star with the radius $\bar{R}$ are represented
as solid black curves. The light curves of stars with NRPs but without
considering the temperature variation over their surface are shown
with dashed black curves. The light curves from NRPs with both
variable radius and variable temperature over their surface in
standard filters BVRI are plotted using blue, dark-green, orange,
and red solid curves.  Their residuals are shown in the bottom
panels. These residuals are the relative difference with respect
to the solid black curves, defined as $\delta A_{o}/A=(A_{o}-A)/A$,
where $A$ is the magnification factor of the corresponding source
star with no pulsational effect. For each panel the inset displays
the projected appearance of the source star at the start time of
the light curve, along with the lens trajectory projected into the
source plane.  The dimensions of the inset are $2\rho_{\star}$ by
$2\rho_{\star}$. Hence, the lens trajectory sometimes falls outside
of the inset and so may not be displayed. The relevant parameters
used to make these microlensing light curves are reported in Table
\ref{table}.  In this table $\xi$ is the angle between the lens
trajectory and the $y_{o}$-axis (horizontal axis on the source
surface). Regarding these light curves, some general points are
listed below.

\begin{enumerate}[label= \arabic*., leftmargin=0.1cm]  

\item  When the lens is far from the source star so that the finite
source effect is negligible, the microlensing light curves with
pulsation arise simply from multiplication of the point lens
magnification function with the intrinsic luminosity curve of the star
undergoing NRPs as if it were a point source. This is the expected
limiting case, and for illustration, two example light curves 
are given in Figures \ref{figc} and \ref{fige2}. \\

\item When the lens impact parameter is small and the finite source
size affects the net magnification factor for the star (i.e., $u
\sim \rho_{\star}$), the difference between the simulated light
curve and the simple microlensing light curve due to a non-pulsating
source star is not simply the intrinsic NRP light curve.  Examples
are given in Figures \ref{figb} and \ref{fige}.  Under these
conditions, the parameters $u_{0}$, $t_{0}$ will be misinterpreted
if finite source effects are not considered.  Imagine that if the
lens at the time of closest approach were to pass across a patch
of the source that happens to have a low effective temperature.
The peak of the light curve may actually decrease despite being at
minimum impact parameter (see \ref{figa2}). \\

\item  For high-magnification microlensing events of NRP stars with
modes $l\geq 2$, the light curves in the different filters show
distinctions from one another during the transit of the source star
by the lens (see Figs. \ref{figd} and \ref{fige}).  Differences
between the filter bands are especially pronounced when the lens
moves over a zone with a higher temperature difference with respect
to $\bar{T}$. For modes with $l=0,~1$, the magnification factor
depends is always filter-dependent, not just while the lens
transits the
source surface (see Fig.~\ref{figa}),
because for these modes the intrinsic luminosity curves of the
source star vary in different filters (e.g., Fig.~\ref{la}).
\\

\item According to point (1) from subsection~\ref{reshap}, for
the modes with Legendre functions proportional to $\cos^{n}\theta_{\star}$,
and for a pole-on view, the star is always circular in projection
with radius $\bar{R}$ versus time.  For these modes, the dashed
black residuals are zero versus time (see Fig.\ref{figb}; since the inclination angle for this plot is not exactly $90~$deg, a pulsation with small amplitude in the dashed black residual is generated). It means
that any deviation in their light curves are due to changes in
the stellar temperature which will be highlighted when the lens
is passing over or close to the stellar surface,  i.e, $u_{0}\sim
\rho_{\star}$. Hence, the amplitude of the variation in the source
radius, $\delta_{\rm R}$, is not measurable even if the lens is
crossing the source surface in these modes. \\

\item For the modes with $m=0$, and when the star's polar axis is
towards the observer ($i=90~$deg), isothermal zones on the source
star are rings concentric to the polar axis. For these modes the circular
symmetry of the source surface is preserved. 
Noting that the phase of the stellar pulsation is
$\omega(t_{0}-t_{\rm p}) + \phi_{0}$,
any asymmetry in the light curve with time of the
closest approach, at the time $t_{0}$, indicates that the pulsation phase
is not at a multiple of $\pi$. 
For the modes with $(l,m)=(1,0)$ and a pole-on view, the effect
of source star pulsation for the microlensing light curve
is similar to the effect of limb-darkening.  \\

\item For modes with $m \neq 0$, and when the star's polar axis
is towards the observer, the time evolution of the NRPs does not
change the temperature profile, but it does rotates the source
surface around the $x_{o}-$axis (pole axis). These stars as seen by the
observer have luminosity profiles that are constant with time and
could errantly be assumed to be non-pulsating stars.  However, when
subject to microlensing, the light curves (especially for a transit
event) display interesting deviations when the lens passes over
the projected stellar limb.  Although the resulting light curve may
be similar to a simple light curve from a non-pulsating source star,
the peak magnification is filter-dependent and chromatic features
appear (see, e.g., \ref{figf}, \ref{figb2}).  The colour dependence
of the magnification factor during transit and for high magnification
microlensing events reveals the variable nature of the source star.
Hence, some stars thought to be constant sources may in reality
be variable, even though their flux is constant for
certain viewing perspectives. The lensing effect can reveal 
their variable nature. \\

\item  From point (5) of subsection (\ref{lumin}),
the region over the stellar equator 
has the largest impact on the intrinsic
luminosity of the source star. Therefore, if the related Legendre
polynomial maximizes for these points, the observer received the
stellar luminosity with the highest amplitude. Accordingly, when
the lens is passing close and parallel with the stellar equator
projected on the sky plane, the deviation in the light curve varies in different filters. 
Because the pulsation in the stellar temperature for these points have the largest amplitude (see Figure
\ref{figd2}).  \\

\item In subsection \ref{reshap}, and with point (3) above, we
notice that the largest amplitude of the reshape factor for the
projected source star happens when the maximum of the related
Legendre polynomial occurs at the stellar poles. Accordingly, when
the lens trajectory comes near the stellar poles projected
onto the sky plane, the deviation in the light curve is due to 
pulsation of the stellar radius rather than the varying
stellar temperature. \\

\item For some NRPs, the brightness centre does not correspond to
the coordinate centre. As an example, consider the mode $(l,m)=(1,0)$. 
When the polar axis is in the plane of the sky, the brightness centre is 
actually displaced in the vertical direction by

\begin{eqnarray}
\Delta_{c}= \cos(i)  \left\{ \delta_{\rm R} \cos [\omega(t-t_{\rm p})] -\bar{R} \frac{\delta_{\rm T}}{\bar{T}} \sin [\omega(t-t_{\rm p})] \right\},
\end{eqnarray}

\noindent where subscript ``c'' refers to centre.  This brightness
centre rotates about the sky plane during the stellar pulsation. For $m\neq0$ if $i=90~$deg, the brightness centre moves around a circle, whereas
for $i=0~$deg, the centre will oscillate along a horizontal line
passing through the coordinate centre. This effect maps into the light
curve as a peak magnification that is temporally offset with respect
to the time of closest approach between the lens and the geometric
centre of the source star. \\

\end{enumerate}

In the following, we explore obervational predictions for
several different pulsation modes with
$l=0,~1,~2,~3,~4,~5$ and $m\in [0, l]$.  For each mode, we mention
the points regarding the source star shape, its area and the
luminosity in different inclination angles. These points somewhat
help  when we aim to discern the pulsation mode of the source star
from the resulted microlensing light curves.

\begin{enumerate}[label= \arabic*), leftmargin=0.1cm] 

\item[$\ast$] $(l,m)=(0,0)$: In this mode the star is actually
radially pulsating for which the luminosity of the source stars is
filter-dependent (see Paper~I for more details). Microlensing
light curves show chromatic effects. The range of variations is
larger in filters at shorter wavelengths. \\

\item[$\ast$] $(l,m)=(1,0)$: In this mode, the source radii in the
direction of the $z_{o}$-axis are $R= \bar{R} \pm \delta_{R} \cos
i \cos [\omega (t-t_{p})]$ and in the direction of the $y_{o}$-axis,
source radii are $\pm \bar{R}$. The source surface has no symmetry
with respect to the horizontal axis, so the brightness centre does
not coincide with the coordinate centre (unless $i=90~$deg). \\


\begin{figure*}
	\centering
	\subfigure[]{\includegraphics[angle=0,width=0.49\textwidth,clip=0]{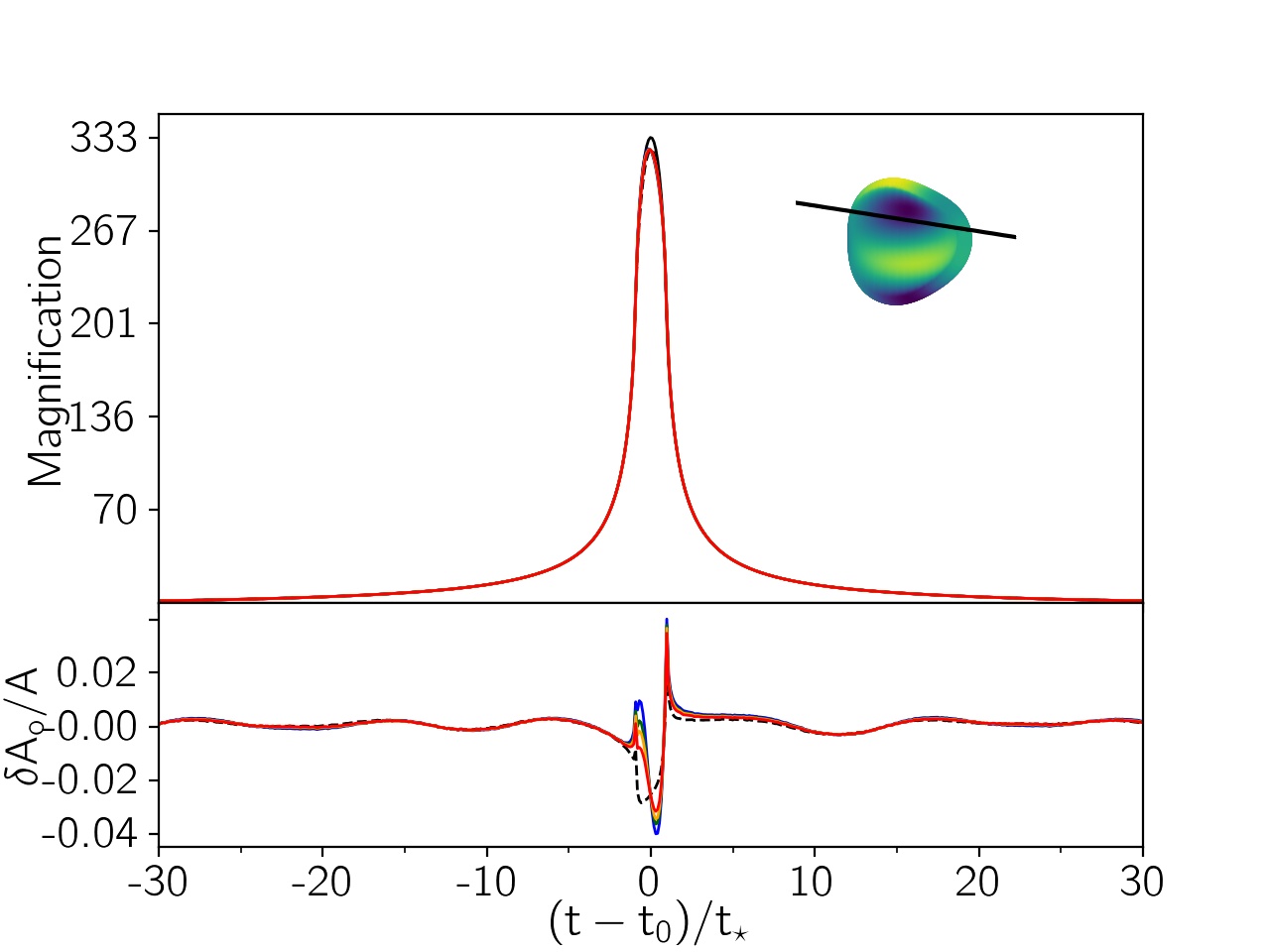}\label{figa2}}
	\subfigure[]{\includegraphics[angle=0,width=0.49\textwidth,clip=0]{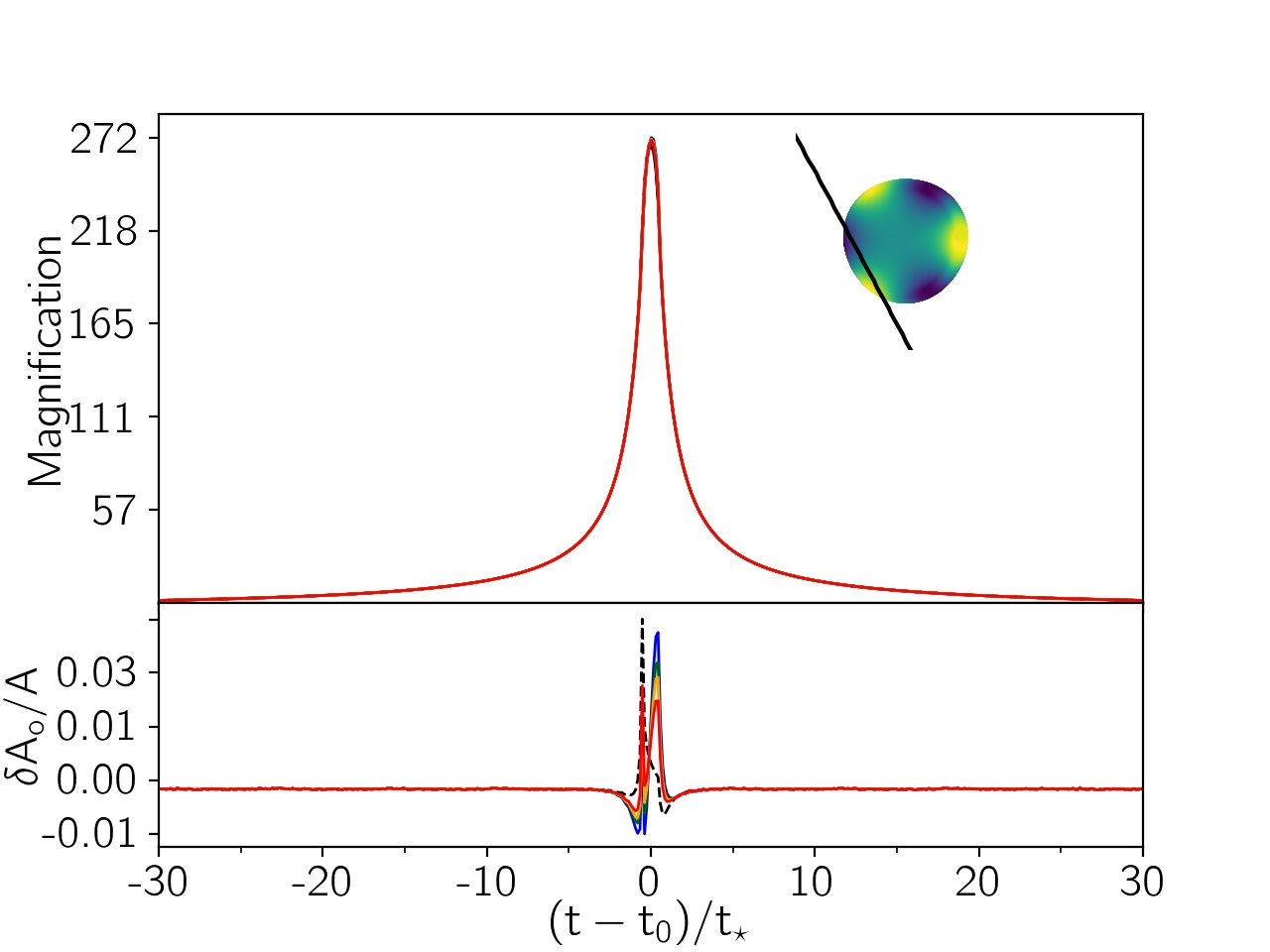}\label{figb2}}
	\subfigure[]{\includegraphics[angle=0,width=0.49\textwidth,clip=0]{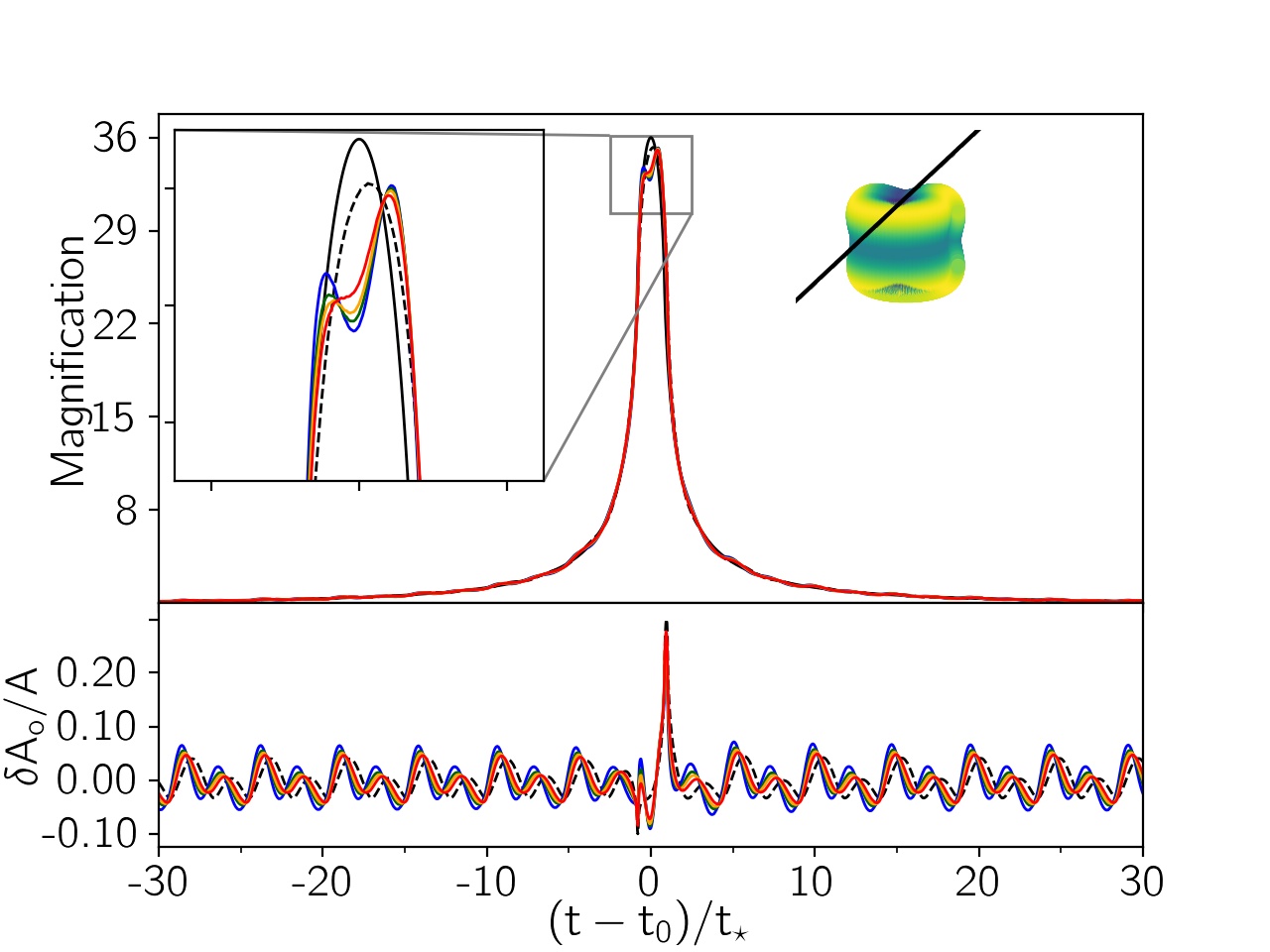}\label{figc2}}
	\subfigure[]{\includegraphics[angle=0,width=0.49\textwidth,clip=0]{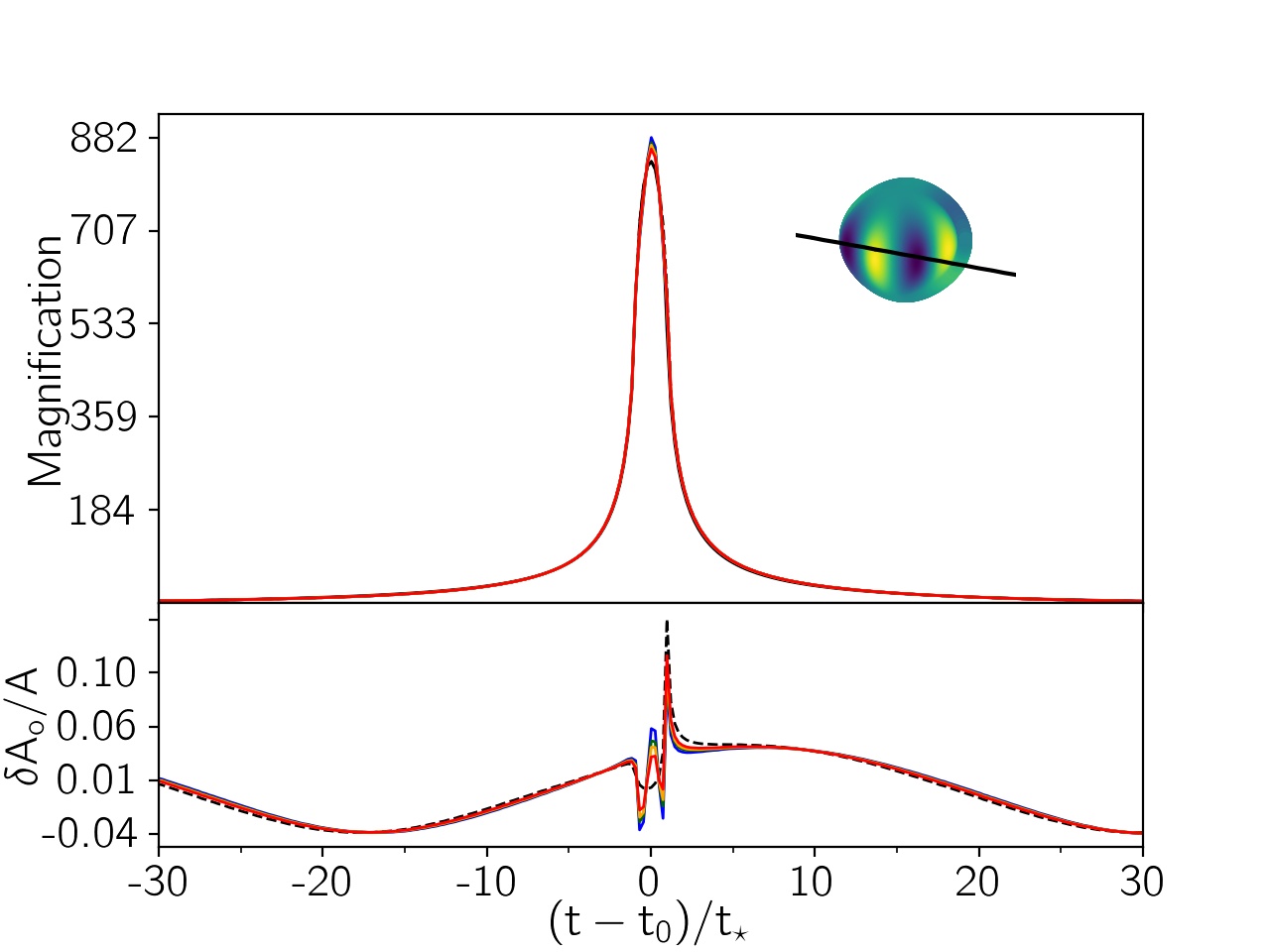}\label{figd2}}
	\subfigure[]{\includegraphics[angle=0,width=0.49\textwidth,clip=0]{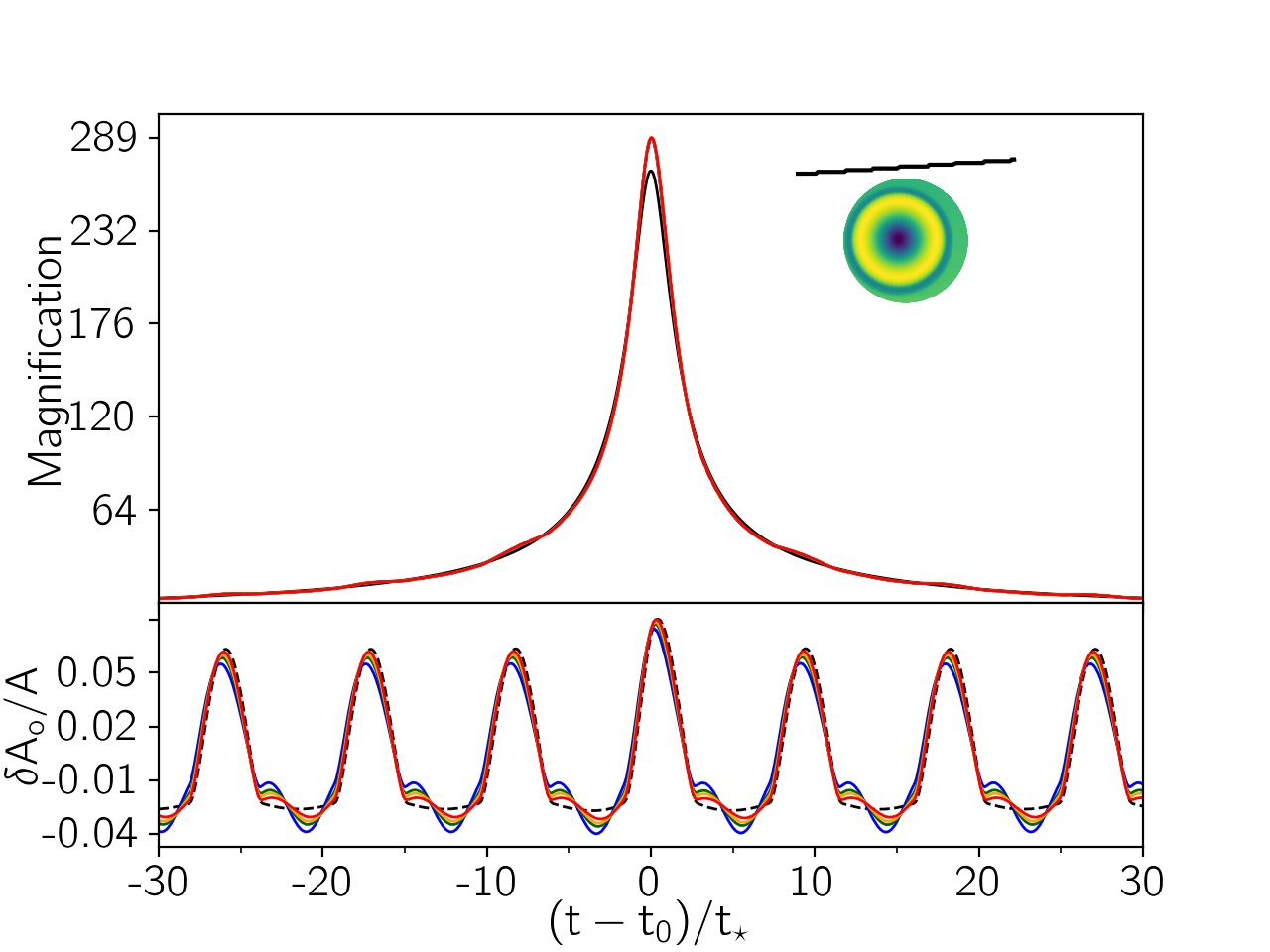}\label{fige2}}
	\subfigure[]{\includegraphics[angle=0,width=0.49\textwidth,clip=0]{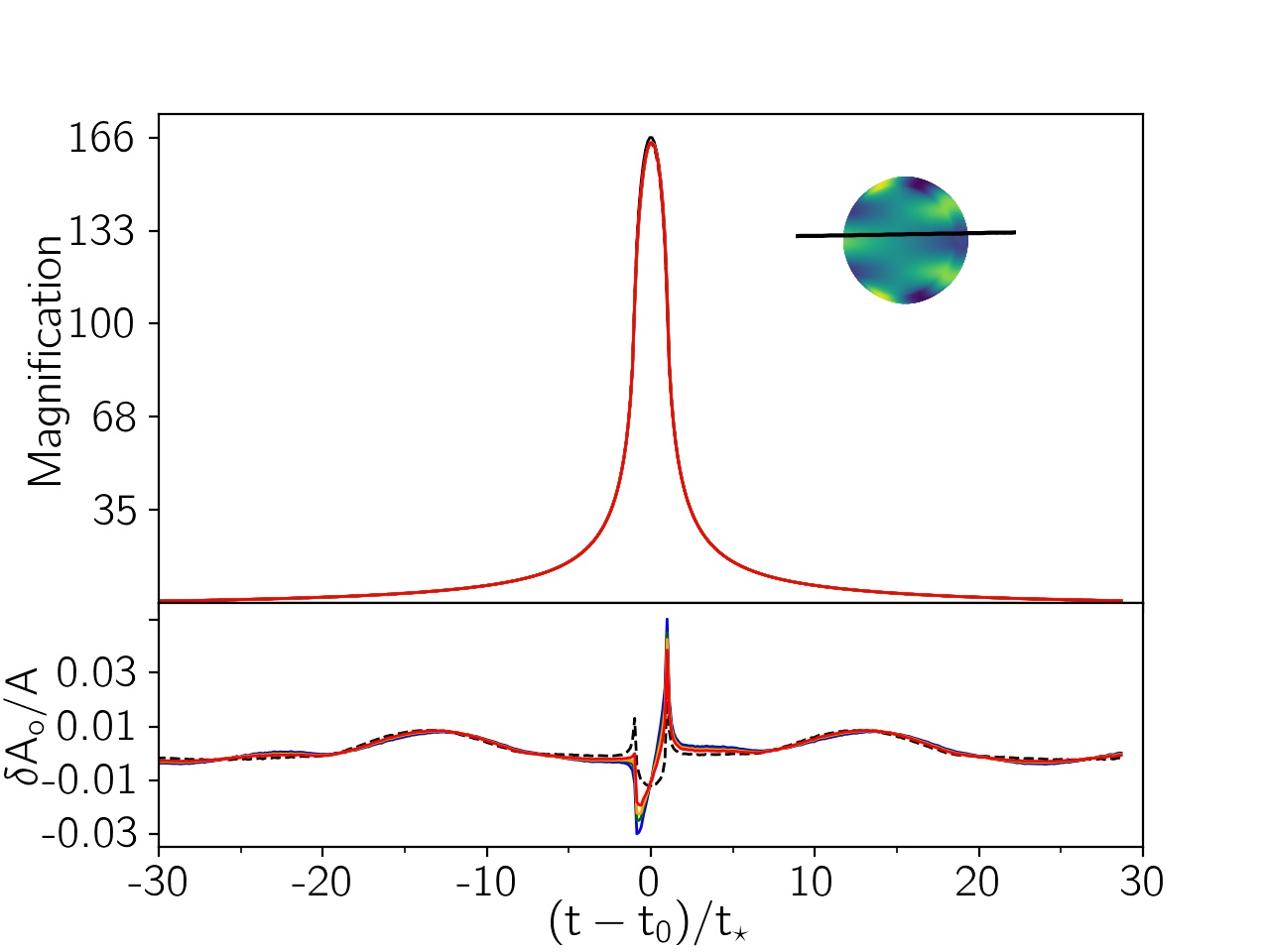}\label{figf2}}
	\caption{Several examples of single microlensing light
	curves from NRPs with different modes, using the same style as
	in Fig.~\ref{light1}.}\label{light2}
\end{figure*}

\item[$\ast$] $(l,m)=(2,0)$: In this mode the source star takes on
the general appearance of an ellipsoid (dipole), with the ratio of
its axes given by 

$$\frac{\bar{R}+\delta_{R} [3 \cos^{2} (i)-1]\cos[\omega (t-t_{p})] }{\bar{R}- \delta_{R} \cos [\omega (t-t_{p})
]}.$$ 

\noindent By increasing the inclination angle, the source shape tends
to a circle, so that when the stellar pole is towards the observer
($i=90$), the source is completely circular but with variable radius
(similar to a radially pulsating star).  When $i=54~$deg (for which
$P^{2}_{0}(i)=0$), the projected source star is also a circle at
fixed radius $\bar{R}$. Finally, the reshape factor is
maximum for the inclination angle $90$ deg. Hence, the deviation in
the light curve due to pulsation in the source radius is different
from deviation due to the stellar temperature (see Figure
\ref{figc}). \\

\begin{table*}
	\centering
	\caption{Table contains the parameters used to make microlensing light curves shown in Figures (\ref{light1}), (\ref{light2}) and (\ref{light3}).}\label{table}
	\begin{tabular}{cccccccccccccc}
		\hline
		& $\delta_{R}$ & $\delta_{T}$  & $P$ & $i$ &  $l$ & $m$ & $u_{\rm{0}}$ & $t_{\rm E}$& $\rho_{\star}$ & $\xi$ & $T$ & $d$ & $q$\\
		& $(\bar{R})$ & $\rm{(K)}$  & $\rm{(day)}$ & $\rm{(deg)}$ &  $~$  & $~$ & $(\rho_{\star})$ & $\rm{(day)}$ & $~$ & $\rm{(deg)}$ & $\rm{(K)}$ & $\rm{(R_{E})}$ & $~$   \\
		\hline
		\ref{figa} &  $0.35$  &  $400$  &  $1.6$  &  $10.2$  &  $1$  &  $1$  &  $1.49$  &  $28.0$  &  $0.011$  &  $273.1$  &  $5727$ &  $-$ &   $-$ \\
		\ref{figb}&  $0.35$  &  $400$  &  $1.8$  &  $88.9$  &  $2$  &  $1$  &  $1.97$  &  $24.0$  &  $0.006$  &  $212.5$  &  $6918$&  $-$ &  $-$\\
		\ref{figc}&  $0.25$  &  $482$  &  $3.7$  &  $32.8$  &  $2$  &  $0$  &  $11.21$  &  $6.3$  &  $0.031$  &  $52.6$  &  $4365$ &  $-$ &  $-$ \\
		\ref{figd}&  $0.30$  &  $463$  &  $0.7$  &  $54.8$  &  $3$  &  $2$  &  $0.13$  &  $29.2$  &  $0.005$  &  $4.3$  &  $4897$ &  $-$ &  $-$\\
		\ref{fige}&  $0.16$  &  $406$  &  $3.8$  &  $77.6$  &  $2$  &  $2$  &  $0.04$  &  $76.5$  &  $0.001$  &  $66.8$  &  $6025$ &  $-$ &  $-$\\
		\ref{figf}&  $0.20$  &  $315$  &  $1.5$  &  $89.4$  &  $3$  &  $0$  &  $0.51$  &  $30.1$  &  $0.002$  &  $26.6$  &  $5211$ &  $-$&  $-$\\
		\ref{figa2}&  $0.24$  &  $437$  &  $2.8$  &  $25.6$  &  $3$  &  $1$  &  $0.37$  &  $21.8$  &  $0.006$  &  $351.0$  &  $5559$ &  $-$ &  $-$\\
		\ref{figb2}&  $0.05$  &  $445$  &  $1.9$  &  $89.2$  &  $3$  &  $3$  &  $0.87$  &  $11.8$  &  $0.0058$  &  $118.2$  &  $5675$ &  $-$ &  $-$\\
		\ref{figc2}&  $0.38$  &  $567$  &  $6.1$  &  $15.7$  &  $4$  &  $0$  &  $0.56$  &  $25.3$  &  $0.0497$  &  $43.1$  &  $3357$ &  $-$ &  $-$ \\
		\ref{figd2}&  $0.18$  &  $581$  &  $6.2$  &  $22.6$  &  $4$  &  $4$  &  $0.27$  &  $55.8$  &  $0.0023$  &  $169.5$  &  $5495$ &  $-$ &  $-$\\
		\ref{fige2}&  $0.22$  &  $559$  &  $0.9$  &  $88.0$  &  $5$  &  $0$  &  $1.34$  &  $32.3$  &  $0.0031$  &  $3.7$  &  $4920$ &  $-$ &  $-$\\
		\ref{figf2}&  $0.10$  &  $471$  &  $1.9$  &  $0.7$  &  $5$  &  $1$  &  $0.10$  &  $6.1$  &  $0.0119$  &  $1.0$  &  $5727$ &  $-$&  $-$\\
		\ref{figa3}  &  $0.32$  &  $325$  &  $1.8$  &  $34.3$  &  $1$  &  $1$  &  $0.024$  &  $12.5$  &  $0.0138$  &  $52.6$  &  $4539$  &  $0.74$  &  $0.81$ \\
		\ref{figb3} &  $0.25$  &  $350$  &  $1.8$  &  $5.0$  &  $2$  &  $1$  &  $0.24$  &  $20.5$  &  $0.0011$  &  $15.8$  &  $4405$  &  $0.89$  &  $0.56$\\
         \ref{figc3}&  $0.31$  &  $313$  &  $1.8$  &  $3.3$  &  $3$  &  $1$  &  $0.21$  &  $8.7$  &  $0.0103$  &  $-0.8$  &  $4666$  &  $1.04$  &  $0.83$\\
        \ref{figd3} &  $0.32$  &  $315$  &  $2.0$  &  $89.0$  &  $5$  &  $1$  &  $0.19$  &  $13.6$  &  $0.0358$  &  $145.1$  &  $4570$  &  $1.10$  &  $0.94$\\
		\hline
	\end{tabular}
\end{table*}
\begin{figure*}
	\centering
	\subfigure[]{\includegraphics[angle=0,width=0.49\textwidth,clip=0]{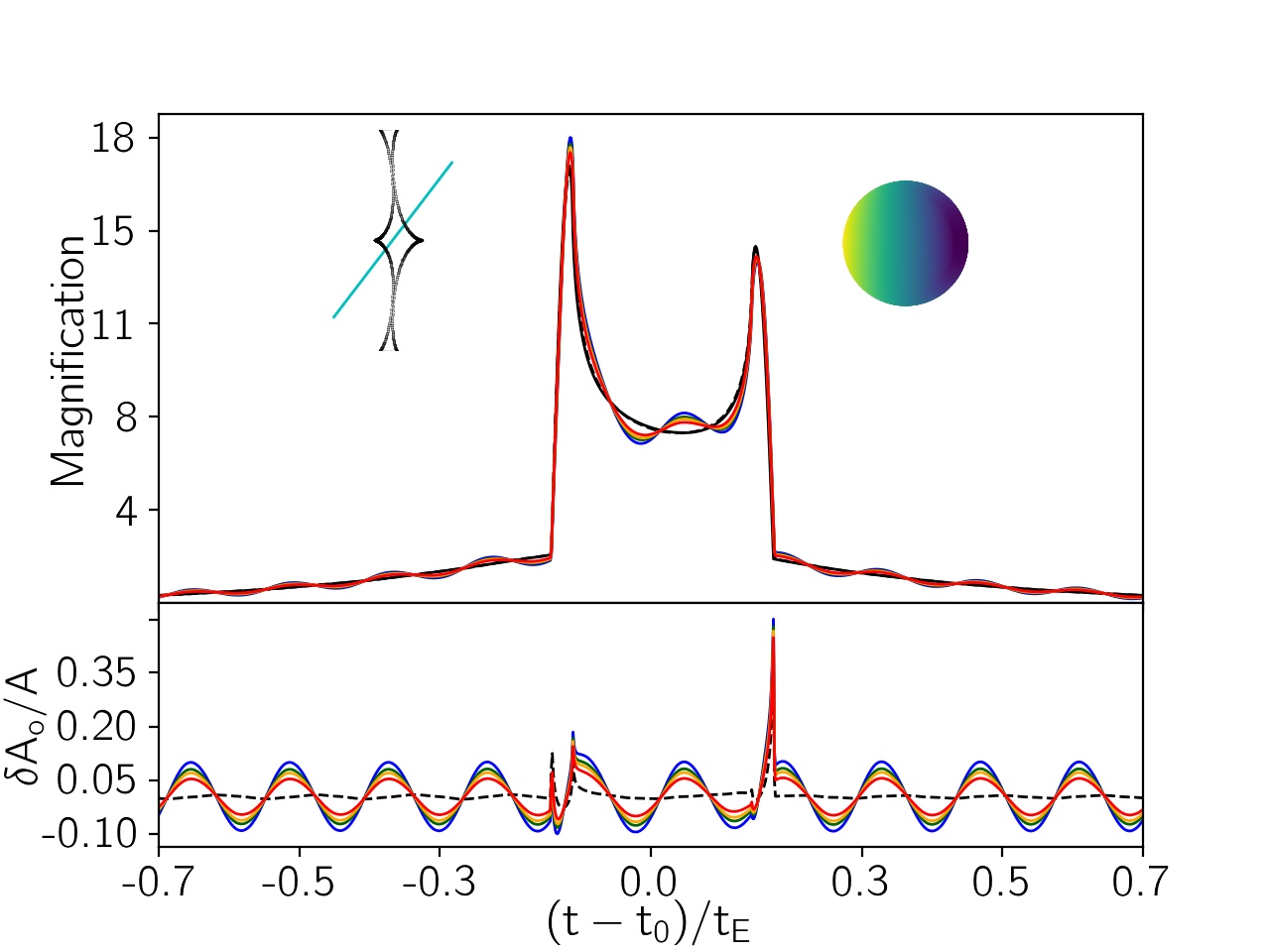}\label{figa3}}
	\subfigure[]{\includegraphics[angle=0,width=0.49\textwidth,clip=0]{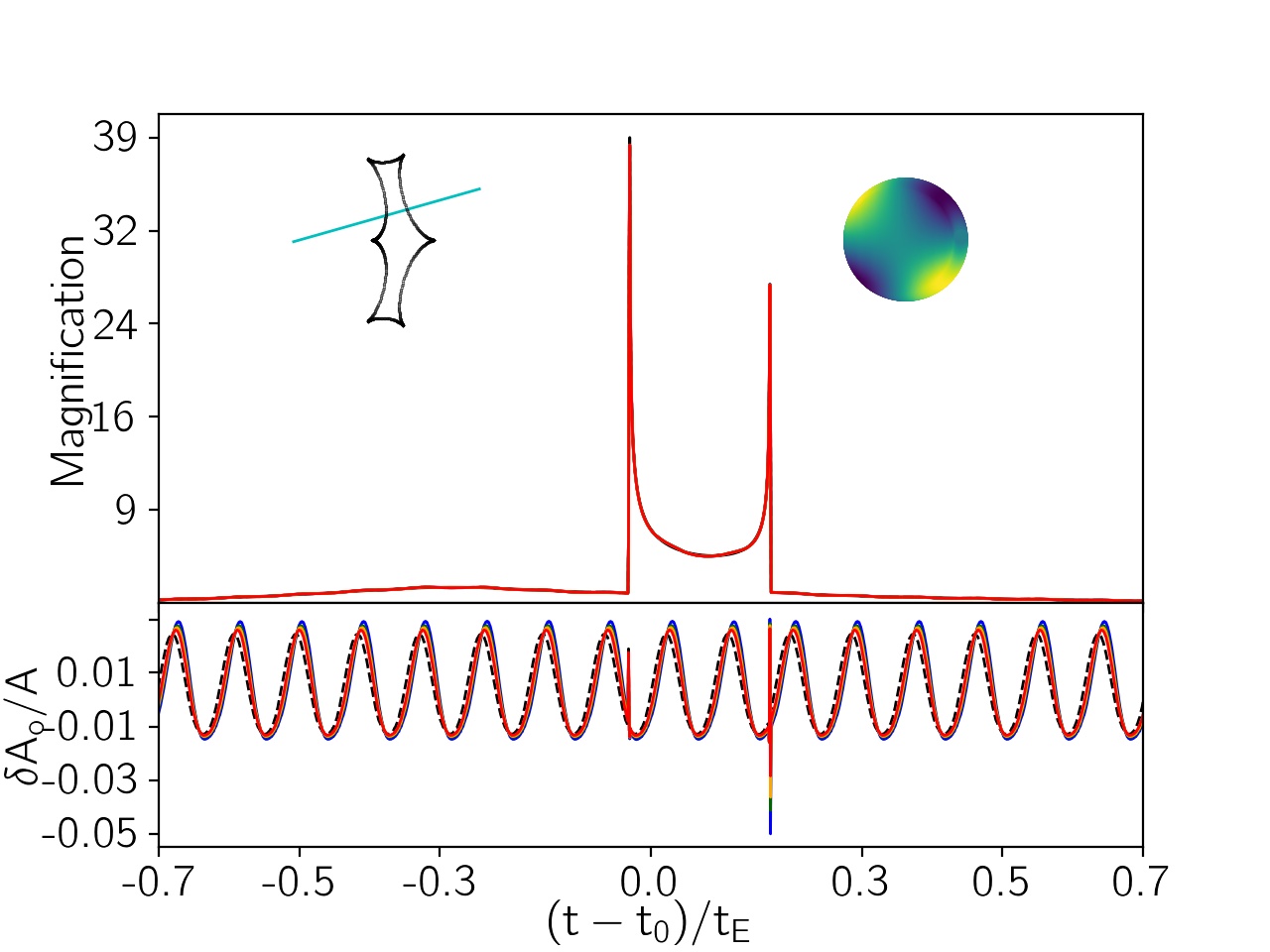}\label{figb3}}
	\subfigure[]{\includegraphics[angle=0,width=0.49\textwidth,clip=0]{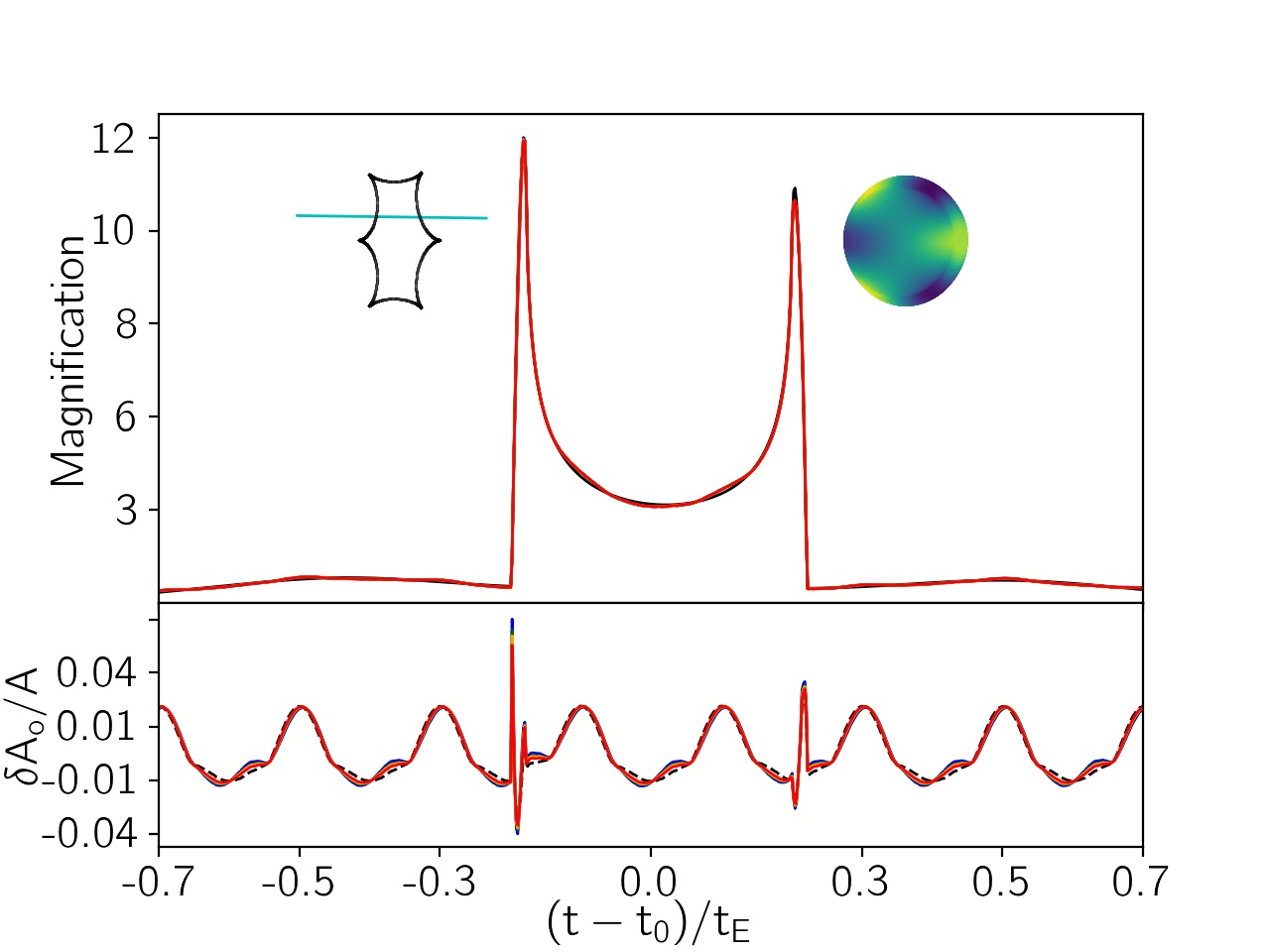}\label{figc3}}
	\subfigure[]{\includegraphics[angle=0,width=0.49\textwidth,clip=0]{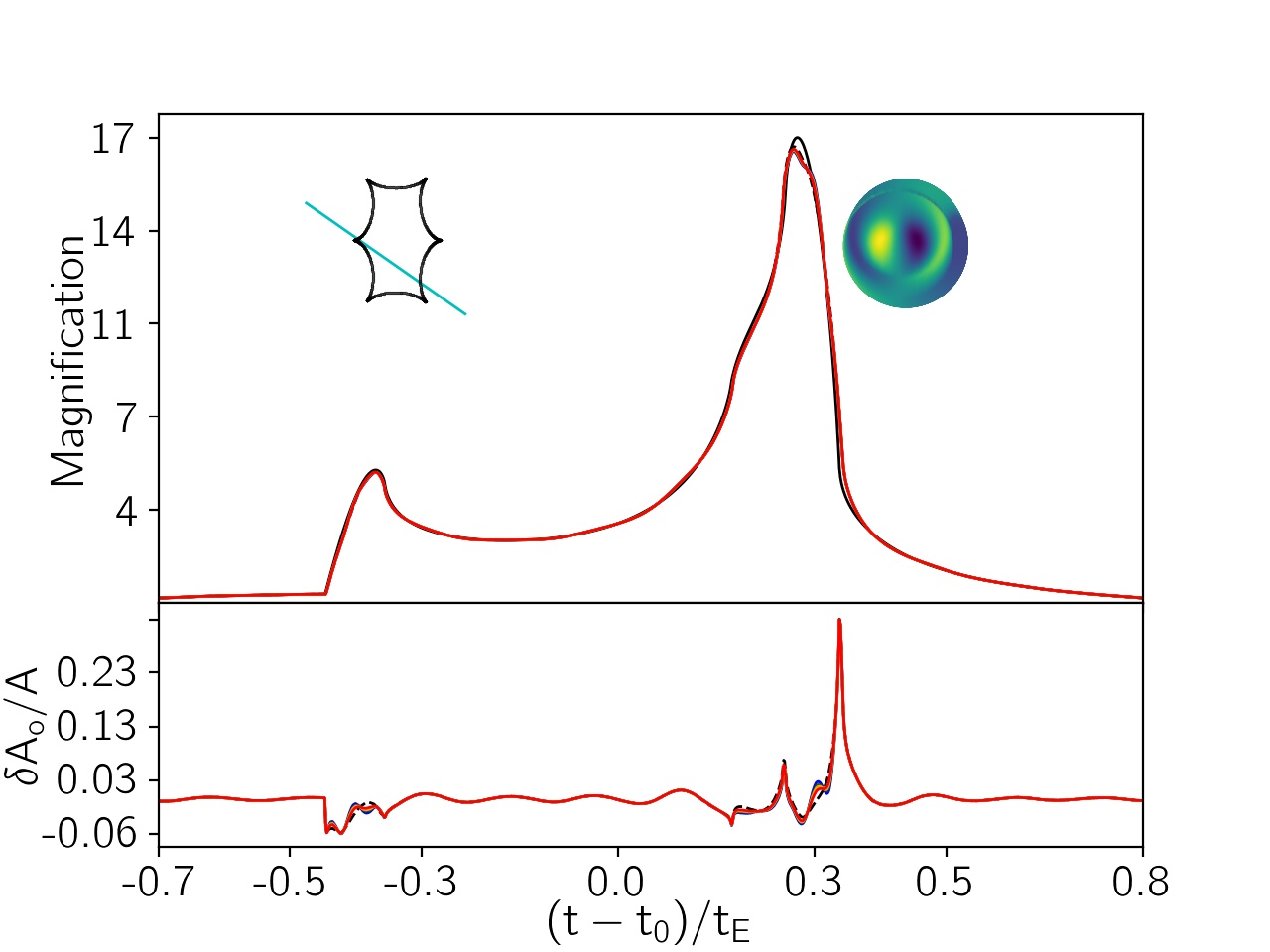}\label{figd3}}
	\caption{Four examples of binary microlensing light curves from NRPs with different modes.  In these plots and in their left-hand insets, the caustic curves (black)  and the source trajectories (cyan)  are shown. The figure otherwise adopts the same style as Fig.~\ref{light1}.}\label{light3}
\end{figure*}

\item[$\ast$] $(l,m)=(2,1)$: For this mode, at
$i\simeq 45~$deg, the amplitudes of the reshape factor and the stellar
luminosity both maximize. When $i=90~$deg,  the source star is
a circle with constant radius $\bar{R}$. In this case the pulsation
effect on the microlensing light curve is very small, unless the
lens is passing close to the source surface. For this mode the
deviation due to transiting over the source surface is chromatic.  \\

\item[$\ast$] $(l,m)=(2,2)$:  The Legendre polynomial for this mode
is $P^{2}_{2}= \sin^{2} \theta_{\star}$. The maximum amplitudes of
the stellar luminosity occurs at $i=0$ deg and the maximum amplitude
of the reshape factor occurs at $i=90$ deg. Hence, the pulsation
effect on the microlensing light curve when the source pole is
towards the observer is very small, unless the lens is crossing the
source surface (c.f., Fig.~\ref{fige}). \\

\item[$\ast$] $(l,m)= (3,0)$: Here $P^{3}_{0}=5 \cos^{3} \theta_{\star}-3
\cos \theta_{\star}$. In this mode the source star around the polar
axis has a conical shape. The maximum reshape factor occurs when
$i=0~$deg when the polar axis is in the plane of the sky (in our
formalism the vertical axis). When $i=90~$deg, although the reshape
factor is zero for all times, the domain of the stellar luminosity
maximizes. The luminosity curves of this mode can be found in
\ref{lb}. During some part of the pulsational phase, differences
between the filter passbands are greatest.  Lensing will increase
this difference further, and one can discern the pulsation mode
according to these coloured features.  An example light curve is
shown in Figure \ref{figf}. \\

\item[$\ast$] $(l,m)=(3,1)$:  The Legendre function is $P^{3}_{1}=
\sin \theta_{\star} \, (5 \cos^{2} \theta_{\star}-1)$. The maximum
reshape factor happens when $i=31, 90~$deg for which $P^{3}_{1}(\cos
i)$ is maximized. We note that for $i\sim 0~$deg, the stellar shape is
conical around the $y_{\star}-$axis. The maximum range in the luminosity
occurs when $i=0,59~$deg. The important points regarding this mode
is that the pulsation in the stellar luminosity is small (the
amplitude of the luminosity is $0.015$ for $\delta_{\rm T}=450~$K
and $\delta_{\rm R}=0.35\bar{R}$). For instance, in Figure \ref{figa2}
significant deviation in the light curve occurs only while transiting
the source surface which is $\sim0.03\%$, although $\delta_{\rm
T}=437~$K.\\

\item[$\ast$] $(l,m)=(3,2)$: For this mode $P^{3}_{2}= \sin^{2}
\theta_{\star} \cos \theta_{\star}$.  Its maximum is achieved for
$\theta_{\star}=0, 55~$deg. Hence, the maximum range in luminosity
occurs when $i\sim 35~$deg. When the pole of the source star is
towards the observer, the star is circular with radius $\bar{R}$.
Similar to the two previous modes, the source has a conical
shape except when $i=90~$deg.  \\

\item[$\ast$] $(l,m)=(3,3)$: Here $P^{3}_{3}=\sin^{3}\theta_{\star}$.
When $i=0~$deg, the domain of the pulsation in the stellar luminosity
maximizes. The maximum reshape factor occurs when the stellar pole
is towards the observer and the stellar shape for this inclination
angle is similar to a triangle (because of the factor $\cos 3
\phi_{\star}$).  When the lens is crossing the source surface, a
substantial asymmetric  deviation in the light curve appears because of the
pulsation in the stellar radius, see, e.g.,  Figure \ref{figb2}.\\

\item[$\ast$] $(l,m)=(4,0)$:  Here $P_{0}^{4}=  35
\cos^{4}\theta_{\star}-30 \cos^{2}\theta_{\star}+3$.  The source
surface when its pole is on the sky plane appears as a diamond (or
square), whereas for $i \sim 90~$deg, the source star is circular
whose radius pulsates over time while the temperature pattern over
the source surface is organized in rings. The amplitude of the
stellar luminosity is large and maximizes when the stellar pole is
towards the observer.  When the lens is goes across the pole a demagnification happens in the light curve and makes the
light curve differ from the simple Pczy\'nski light curve (similar
to binary or planetary microlensing ones); see Figure \ref{figc2}. We note that time variable colour during the
transit of the star by the lens can produce chromatic perturbations
to reveal stellar brightness anomalies. Binary or
planetary microlensing may have similar deviations in their light curves, but they do not make chromatic effects.\\

\item[$\ast$] $(l,m)=(4,1)$: The Legandre function is $P_{1}^{4}=
\sin \theta_{\star} \cos \theta_{\star} (7 \cos^{2} \theta_{\star}-3)$.
When $i=90~$deg, the source star is circular with radius
$\bar{R}$.  The highest reshape factor happens when
$i=0~$deg, and
the source star as seen by the observer appears as a diamond. When $i=66$
the largest amplitude for the stellar luminosity is achieved. \\

\item[$\ast$] $(l,m)=(4,2)$: In this mode the maximum amplitude for
the stellar luminosity is achieved at $i=49~$deg. 
The largest
reshape factor occurs when $i=90~$deg and the source star shape is
dipolar, but for other values of the inclination angle, the source
shape is like a diamond.  \\

\item[$\ast$] $(l,m)=(4,3)$:  The maximum of the
Legandre polynomial occurs at $\theta_{\star}=60~$deg. Hence, for
$i=60~$deg, the source projected on the sky plane looks like a
diamond with a high reshape factor.  For $\delta_{R}=0.35$ and
$\delta_{\rm T}=450~$K, the maximum luminosity amplitude is 1.1,
which is smaller than the amplitude of the luminosity for the other modes
with $l=4$ which have $m \neq 3$. \\

\item[$\ast$] $(l,m)=(4,4)$: At $i=0~$deg, the source surface is
not symmetric with respect to the vertical axis ($z_{\star}$), and
for this inclination angle, the amplitude of the luminosity curve
maximizes.  While the lens is crossing the source surface, and
specially when it passes horizontal, the light curve will be chromatic
even though the intrinsic luminosity curve is not filter-dependent.
One example light curve is shown in Figure \ref{figd2}.  When the
stellar pole is towards the observer, its intrinsic luminosity does
not change with time (although its reshape factor maximizes and the
source surface looks diamond).  Without the effect of lensing, this
star could not be discerned as a variable. \\

\item[$\ast$] $(l,m)=(5,0)$:  For this mode,  when the inclination
angle is small, the projected surface of the source star on the
sky is similar to a pentagon, and by increasing the inclination
angle, the source surface becomes more symmetric so that when
$i=90~$deg, the source star is a circle with radius $\bar{R}$.
However, the stellar pole is towards the observer, the highest
amplitude of its intrinsic luminosity curve occurs.  The source
star in this mode and with $i=90$  is similar to the
source star with the mode $(l,m)=(3,0)$ and $i=90~$deg.  For both
modes, the coloured features appear at the downstream of the luminosity
curve.  In Figure \ref{fige2}, one example microlensing light curve
from an NRP in this mode is shown. The residual light curve is
similar to the cyan curve in Figure \ref{lb}. \\

\item[$\ast$] $(l,m)=(5,1)$:  Here, the source star (similar
to the previous mode) looks like a pentagon when the stellar pole is
in the sky plane.  However for this mode the stellar pole is
horizontal axis. The Legendre polynomial in this mode will be zero
at $\theta_{\star}=73~$deg which results a very small reshape factor
when $i=73~$deg.  The maximum of the Legendre polynomial happens
when $\theta_{\star}=19.4~$deg, such that the amplitude of the
luminosity curve maximizes when
$i=70.6~$deg. Generally, for the modes with $l=5$ and when $m>0$,
the integrated luminosity of the source star pulsates with small
amplitude $<0.1L(\bar{T}, \bar{R})$. Hence, just when the lens is
crossing the source surface, the light curve will deviate from
a simple Paszy\'nski light curve; see, e.g., Figure \ref{figf2}.  \\

\item[$\ast$] $(l,m)=(5,2)$: Similar to the two previous
modes, the source star 
appears as a pentagon when the stellar pole
is in the sky plane, and appears as a circle of radius $\bar{R}$ when
its pole is towards the observer. The maximum of the Legendre function
for this mode occurs at $\theta_{\star}=32~$deg, hence the maximum
amplitude of the luminosity function happens at $i=58~$deg. \\

\item[$\ast$] $(l,m)=(5,3)$: Since the maximum of the Legendre
polynomial in this mode corresponds to $\theta_{\star}=46~$deg, the
maximum amplitude of the stellar luminosity curve is achieved at
$i=44~$deg. The smallest amplitude of the reshape factor for this
mode happens at $i=70~$deg, where the Legendre function vanishes.
\\

\item[$\ast$] $(l,m)=(5,4)$:  The largest amplitude of the luminosity
and the reshape factor happen at $i=27, 63~$deg, respectively.

\item[$\ast$] $(l,m)=(5,5)$:  The maximum luminosity happens
when the stellar pole is on the sky plane, although in this position the
reshape factor is very small.  Also when the stellar pole is towards
the observer, the source star is like as pentagon (because of the
factor $\cos (5\phi)$) but its luminosity does not change. We note that the
amplitude of the luminosity in this mode is the smallest of
the modes with $l=5$.  
\end{enumerate}

\section{Binary Microlensing of Non-radially pulsating stars}\label{binary} 

Around $6\%$ of microlensing light curves have caustic-crossing
features and so-called binary microlensing events
\citep{sumi2013,ogle2019}. These events are more sensitive to
exoplanets than for a single lens, so that most of these events (especially ones with
caustic-crossing features) are fully covered with observational
data taken by survey and follow-up telescopes. In this section, we
simulate the caustic-crossing microlensing events of NRP stars
to examine the properties of the light curves. We aim to investigate
where and how the intrinsic pulsation of the source stars modifies
the microlensing light curves.

In order to calculate the magnified stellar
luminosity of NRP stars lensed by binary microlenses, given by
Equation \ref{lstar}, we evaluate the magnification factor for each
element over the source surface using the well-developed $\rm{RT}$-model
of V.\ Bozza \citep{Bozza2018,Bozza2010,Skowron2012}.  We produce
a large ensemble of caustic-crossing binary microlensing events in
the following way.  We randomly sample the mass ratio of microlenses
in the range $q\in [0, 1]$, as well as their projected distances
on the sky plane normalized to the Einstein radius from the range
$d(R_{\rm E})\in [0.4, 2.0]$. In Figure \ref{light3}, four examples
of binary light curves from NRP stars are presented. Their parameters
can be found in Table~\ref{table}. The characterizations of these
light curves are similar to the ones plotted in Figures~\ref{light1}
and \ref{light2}.  Upper left-hand insets display the caustic curves
(black curves) and the source trajectories (cyan straight lines).
Using the examples of the figure, we offer these summary points.

\begin{itemize}
\item[*] The pulsation amplitudes are more
increased while the source star is inside the caustic curve than
when it is outside (see Fig. \ref{figa3}).  However, the detectability
of the magnified pulsational variability depends on the number of
pulsations occurring when the source is inside the caustic curves.\\

\item[*] Generally, the pulsation of the source star will have three
effects on the magnification peak while transiting the caustic which
are: (i) variation of the stellar colour (Fig.  \ref{figb3}); (ii)
displacement in the peak position; and (iii) variation in the value
of the magnification factor. The variation in the stellar colour
while transiting the caustic line decreases with increasing pulsation
mode, because the temperature contrast over the source surface
decreases with larger $l$ values.  Deviations in the form of
increase-zero-decrease (or decrease-zero-increase) made while
transiting the caustic curve reveals the displacement at the time
when the source brightness centre is on the caustic curve, as seen
for example in the residual curves of Figure \ref{figc3}.  The
amount of the magnification factor at the peak depends on the source
radius according to its value at the time of caustic crossing. \\

\item[*]In the case that the intrinsic pulsation of the
source star is ignored, the two latter effects of NRP
stars on the magnification peak during caustic crossing can lead to
an errant measure of the Einstein crossing time and the parameters of
lens trajectories (the lens impact parameter and the angle of the
lens trajectory with respect to the binary axis). We note that the
first effect (chromatic deviations) is helpful for discerning the
intrinsic pulsation of the source star.\\

\item[*] When the source is relatively large, the
peak of the magnification factor becomes flattened, and pulsation
of the stellar radius changes the detailed shape of the flattened peak (see Fig. \ref{figd3}).

\end{itemize}

Unlike single lens, caustic crossing means makes two high magnification events of the NRP star. Hence, the ability of inferring the $(l,~m)$ modes of the pulsations is higher in binary lensing, because the star is scanned twice (essentially resolved) but at different phases in the pulsation.  However, the parameters of  binary lensing events are more than single ones. So, these events offer a wide diversity of possibilities and capabilities for inferring the stellar properties and the lensing configurations.

\section{Summarize and conclusions}\label{conclu}

Our focus for this paper has been a parameter study for microlensing
of stars undergoing NRPs.  The luminosity variations are characterized
in terms of the stellar radius and surface effective temperature
using harmonic oscillating functions with different $l,m$ values.
We consider a projection angle $i$ between the source pole  and the sky
plane to project the source surface onto the sky plane.
Using this formalism, we simulated microlensing light curves
from NRP stars with different modes and projection
angles both for single lenses and for binary lenses. 

We summary main points first for the single lens case.  When the
lens has a relatively large impact parameter from the NRP star, the
light curve with microlensing is a simple multiplication of the
magnification factor and the stellar luminosity curve. When the
lens is close to the stellar surface or actually has a transit, the
magnification factor completely differs from the simple light curve
of a non-pulsating source star.

When $m \neq 0$ and the star is viewed pole-on, the luminosity from
integration over the source surface is constant with time, so that
although the star displays variability patterns across its surface,
the unresolved star can mimic a constant source.  Lensing effects,
especially when the lens impact parameter is small with $u_{0}
\lesssim \rho_{\star}$, breaks this degeneracy to reveal the intrinsic
surface variability of the source star. During a transit event of a NRP star, the magnification factor is found to be chromatic. 
For non-pulsing stars and in high magnification microlensing events, the chromatic effects can be created when the source star is a red giant with the considerable limb-darkening effect \citep{1998david}.

For some pulsation modes (e.g., $m=1$), the temperature pattern on the
source surface makes the brightness centre shift from the coordinate
centre of the star.  Plus the displacement between brightness centre
and the geometric centre moves with time. This affects the location
of the peak magnification in the light curve.

Some pulsation modes lead to quite similar luminosity curves, e.g.,
$(l,m)=(1,1)$ and $(2,1)$. 
The distinctions in the modes are not recognizable from the normal light
curve.  But during high magnification and transiting microlensing
events, the lensed light curve can distinguish between the modes,
because the different models produce different temperature patterns.\\
    
In binary microlensing events of NRP source
stars, the variation in the stellar colour while the source surface
transits a caustic curve hints at the stellar pulsation
even if its amplitude is too small to produce clear periodic
fluctuations. In the case that the stellar variations are not
discerned, their effects on the magnification peak and its location
can lead to errors of the lensing parameters.

Around $10\%$ of the Galactic Bulge stars are
variable. By considering that annually around $3000$ microlensing
events are identified by survey microlensing groups
\citep{OGLEIII,Sako2008,KMTNet}, the number of microlensing events
due to variable stars should be $\sim 300$ per year. Actually, this
value may be considered a lower limit, since around $30\%$ of
detected microlensing events are due to giant stars, and the
likelihood of variability among the giants is higher than dwarf
stars. Using $137$ microlensing candidates with stellar variability
signatures as a baseline from OGLE-III observations during 2001-2004
\citep{Lukaz2006}, it appears that many of these variable stars are
faint. It is possible that significant deviations only occur when
the magnification factor is high, and their perturbations are likely
mis-interpreted. Statistical and detailed observational issues
involving microlensing events of variable stars will be quantitatively
studied in a subsequent paper.

\section*{Data availability}

Data available on request.

\bibliographystyle{mnras}
\bibliography{ref_nrp}

\begin{thebibliography}{}
\makeatletter
\relax
\def\mn@urlcharsother{\let\do\@makeother \do\$\do\&\do\#\do\^\do\_\do\%\do\~}
\def\mn@doi{\begingroup\mn@urlcharsother \@ifnextchar [ {\mn@doi@}
  {\mn@doi@[]}}
\def\mn@doi@[#1]#2{\def\@tempa{#1}\ifx\@tempa\@empty \href
  {http://dx.doi.org/#2} {doi:#2}\else \href {http://dx.doi.org/#2} {#1}\fi
  \endgroup}
\def\mn@eprint#1#2{\mn@eprint@#1:#2::\@nil}
\def\mn@eprint@arXiv#1{\href {http://arxiv.org/abs/#1} {{\tt arXiv:#1}}}
\def\mn@eprint@dblp#1{\href {http://dblp.uni-trier.de/rec/bibtex/#1.xml}
  {dblp:#1}}
\def\mn@eprint@#1:#2:#3:#4\@nil{\def\@tempa {#1}\def\@tempb {#2}\def\@tempc
  {#3}\ifx \@tempc \@empty \let \@tempc \@tempb \let \@tempb \@tempa \fi \ifx
  \@tempb \@empty \def\@tempb {arXiv}\fi \@ifundefined
  {mn@eprint@\@tempb}{\@tempb:\@tempc}{\expandafter \expandafter \csname
  mn@eprint@\@tempb\endcsname \expandafter{\@tempc}}}

\bibitem[\protect\citeauthoryear{{Alcock} et~al.,}{{Alcock}
  et~al.}{1997}]{alcock1997}
{Alcock} C.,  et~al., 1997, \mn@doi [\apjl] {10.1086/311053}, \href
  {https://ui.adsabs.harvard.edu/abs/1997ApJ...491L..11A} {491, L11}

\bibitem[\protect\citeauthoryear{{Alcock} et~al.,}{{Alcock}
  et~al.}{2000}]{alcock2000}
{Alcock} C.,  et~al., 2000, \mn@doi [\apj] {10.1086/309512}, \href
  {https://ui.adsabs.harvard.edu/abs/2000ApJ...542..281A} {542, 281}

\bibitem[\protect\citeauthoryear{{Assef} et~al.,}{{Assef}
  et~al.}{2006}]{Assef2006}
{Assef} R.~J.,  et~al., 2006, \mn@doi [\apj] {10.1086/506439}, \href
  {https://ui.adsabs.harvard.edu/abs/2006ApJ...649..954A} {649, 954}

\bibitem[\protect\citeauthoryear{{Bonanno} \& {Sereno}}{{Bonanno} \&
  {Sereno}}{2004}]{Bonanno2004}
{Bonanno} A.,  {Sereno} M.,  2004, in {Favata} F.,  {Aigrain} S.,   {Wilson}
  A.,  eds,  ESA Special Publication Vol. 538, Stellar Structure and Habitable
  Planet Finding. pp 281--283

\bibitem[\protect\citeauthoryear{{Bozza}}{{Bozza}}{2010}]{Bozza2010}
{Bozza} V.,  2010, \mn@doi [\mnras] {10.1111/j.1365-2966.2010.17265.x}, \href
  {http://adsabs.harvard.edu/abs/2010MNRAS.408.2188B} {408, 2188}

\bibitem[\protect\citeauthoryear{{Bozza}, {Bachelet}, {Bartoli{\'c}}, {Heintz},
  {Hoag}  \& {Hundertmark}}{{Bozza} et~al.}{2018}]{Bozza2018}
{Bozza} V.,  {Bachelet} E.,  {Bartoli{\'c}} F.,  {Heintz} T.~M.,  {Hoag} A.~R.,
    {Hundertmark} M.,  2018, \mn@doi [\mnras] {10.1093/mnras/sty1791}, \href
  {http://adsabs.harvard.edu/abs/2018MNRAS.479.5157B} {479, 5157}

\bibitem[\protect\citeauthoryear{Carroll \& Ostlie}{Carroll \&
  Ostlie}{1996}]{Carrollbook}
Carroll B.~W.,  Ostlie D.~A.,  1996, {An introduction to modern astrophysics;
  1st ed.}.
Addison-Wesley, Reading, MA, \url {https://cds.cern.ch/record/642519}

\bibitem[\protect\citeauthoryear{{Catelan} \& {Smith}}{{Catelan} \&
  {Smith}}{2015}]{book2015}
{Catelan} M.,  {Smith} H.~A.,  2015, {Pulsating Stars}

\bibitem[\protect\citeauthoryear{{Cowling}}{{Cowling}}{1941}]{cowling1941}
{Cowling} T.~G.,  1941, \mn@doi [\mnras] {10.1093/mnras/101.8.367}, \href
  {https://ui.adsabs.harvard.edu/abs/1941MNRAS.101..367C} {101, 367}

\bibitem[\protect\citeauthoryear{{Cox}}{{Cox}}{1974}]{Cox1974}
{Cox} J.~P.,  1974, \mn@doi [Reports on Progress in Physics]
  {10.1088/0034-4885/37/5/001}, \href
  {https://ui.adsabs.harvard.edu/abs/1974RPPh...37..563C} {37, 563}

\bibitem[\protect\citeauthoryear{{Derekas}, {Kiss}  \& {Bedding}}{{Derekas}
  et~al.}{2007}]{derekas2007}
{Derekas} A.,  {Kiss} L.~L.,   {Bedding} T.~R.,  2007, \mn@doi [\apj]
  {10.1086/517994}, \href
  {https://ui.adsabs.harvard.edu/abs/2007ApJ...663..249D} {663, 249}

\bibitem[\protect\citeauthoryear{{Dominik}}{{Dominik}}{2006}]{Dominik2006}
{Dominik} M.,  2006, \mn@doi [\mnras] {10.1111/j.1365-2966.2006.10004.x}, \href
  {https://ui.adsabs.harvard.edu/abs/2006MNRAS.367..669D} {367, 669}

\bibitem[\protect\citeauthoryear{{Dziembowski}}{{Dziembowski}}{1977}]{Dziem1977}
{Dziembowski} W.,  1977, \actaa, \href
  {https://ui.adsabs.harvard.edu/abs/1977AcA....27..203D} {27, 203}

\bibitem[\protect\citeauthoryear{{Einstein}}{{Einstein}}{1936}]{Einstein1936}
{Einstein} A.,  1936, \mn@doi [Science] {10.1126/science.84.2188.506}, \href
  {http://adsabs.harvard.edu/abs/1936Sci....84..506E} {84, 506}

\bibitem[\protect\citeauthoryear{{Gaudi}}{{Gaudi}}{2012}]{Gaudi2012}
{Gaudi} B.~S.,  2012, \mn@doi [\araa] {10.1146/annurev-astro-081811-125518},
  \href {https://ui.adsabs.harvard.edu/abs/2012ARA&A..50..411G} {50, 411}

\bibitem[\protect\citeauthoryear{{Gaudi} \& {Haiman}}{{Gaudi} \&
  {Haiman}}{2004}]{gaudi2004}
{Gaudi} B.~S.,  {Haiman} Z.,  2004, arXiv e-prints, \href
  {https://ui.adsabs.harvard.edu/abs/2004astro.ph..1035G} {pp
  astro--ph/0401035}

\bibitem[\protect\citeauthoryear{{Graczyk} et~al.,}{{Graczyk}
  et~al.}{2011}]{Graczyk2011}
{Graczyk} D.,  et~al., 2011, \actaa, \href
  {https://ui.adsabs.harvard.edu/abs/2011AcA....61..103G} {61, 103}

\bibitem[\protect\citeauthoryear{{Han}, {Park}, {Kim}  \& {Chang}}{{Han}
  et~al.}{2000}]{han2000}
{Han} C.,  {Park} S.-H.,  {Kim} H.-I.,   {Chang} K.,  2000, \mn@doi [\mnras]
  {10.1046/j.1365-8711.2000.03534.x}, \href
  {https://ui.adsabs.harvard.edu/abs/2000MNRAS.316..665H} {316, 665}

\bibitem[\protect\citeauthoryear{{Hendry}, {Bryce}  \& {Valls-Gabaud}}{{Hendry}
  et~al.}{2002}]{Hendry2002}
{Hendry} M.~A.,  {Bryce} H.~M.,   {Valls-Gabaud} D.,  2002, \mn@doi [\mnras]
  {10.1046/j.1365-8711.2002.05496.x}, \href
  {https://ui.adsabs.harvard.edu/abs/2002MNRAS.335..539H} {335, 539}

\bibitem[\protect\citeauthoryear{{Heyrovsk{\'y}} \& {Loeb}}{{Heyrovsk{\'y}} \&
  {Loeb}}{1997}]{heyrovski1997}
{Heyrovsk{\'y}} D.,  {Loeb} A.,  1997, \mn@doi [\apj] {10.1086/304855}, \href
  {https://ui.adsabs.harvard.edu/abs/1997ApJ...490...38H} {490, 38}

\bibitem[\protect\citeauthoryear{{Heyrovsk{\'y}}, {Sasselov}  \&
  {Loeb}}{{Heyrovsk{\'y}} et~al.}{2000}]{Heyrovski2000}
{Heyrovsk{\'y}} D.,  {Sasselov} D.,   {Loeb} A.,  2000, \mn@doi [\apj]
  {10.1086/317067}, \href
  {https://ui.adsabs.harvard.edu/abs/2000ApJ...543..406H} {543, 406}

\bibitem[\protect\citeauthoryear{{Kahn}}{{Kahn}}{1969}]{Kahn1969}
{Kahn} F.~D.,  1969, \mn@doi [\planss] {10.1016/0032-0633(69)90176-7}, \href
  {https://ui.adsabs.harvard.edu/abs/1969P&SS...17.1563K} {17, 1563}

\bibitem[\protect\citeauthoryear{{Kim} et~al.,}{{Kim} et~al.}{2018}]{KMTNet}
{Kim} D.~J.,  et~al., 2018, \mn@doi [\aj] {10.3847/1538-3881/aaa47b}, \href
  {https://ui.adsabs.harvard.edu/abs/2018AJ....155...76K} {155, 76}

\bibitem[\protect\citeauthoryear{{Li} et~al.,}{{Li} et~al.}{2019}]{Varmicro}
{Li} S.~S.,  et~al., 2019, \mn@doi [\mnras] {10.1093/mnras/stz1873}, \href
  {https://ui.adsabs.harvard.edu/abs/2019MNRAS.488.3308L} {488, 3308}

\bibitem[\protect\citeauthoryear{{Mao} \& {Paczynski}}{{Mao} \&
  {Paczynski}}{1991}]{Mao1991planet}
{Mao} S.,  {Paczynski} B.,  1991, \mn@doi [\apjl] {10.1086/186066}, \href
  {https://ui.adsabs.harvard.edu/abs/1991ApJ...374L..37M} {374, L37}

\bibitem[\protect\citeauthoryear{{Moniez}, {Sajadian}, {Karami}, {Rahvar}  \&
  {Ansari}}{{Moniez} et~al.}{2017}]{moniez}
{Moniez} M.,  {Sajadian} S.,  {Karami} M.,  {Rahvar} S.,   {Ansari} R.,  2017,
  \mn@doi [\aap] {10.1051/0004-6361/201730488}, \href
  {https://ui.adsabs.harvard.edu/abs/2017A&A...604A.124M} {604, A124}

\bibitem[\protect\citeauthoryear{{Mr{\'o}z} et~al.,}{{Mr{\'o}z}
  et~al.}{2019}]{ogle2019}
{Mr{\'o}z} P.,  et~al., 2019, \mn@doi [\apjs] {10.3847/1538-4365/ab426b}, \href
  {https://ui.adsabs.harvard.edu/abs/2019ApJS..244...29M} {244, 29}

\bibitem[\protect\citeauthoryear{{Paczynski}}{{Paczynski}}{1986}]{pac86}
{Paczynski} B.,  1986, \mn@doi [\apj] {10.1086/163919}, \href
  {http://adsabs.harvard.edu/abs/1986ApJ...301..503P} {301, 503}

\bibitem[\protect\citeauthoryear{{Percy}}{{Percy}}{2007}]{book2007}
{Percy} J.~R.,  2007, {Understanding Variable Stars}

\bibitem[\protect\citeauthoryear{{Pietrukowicz} et~al.,}{{Pietrukowicz}
  et~al.}{2013}]{OGLE2013}
{Pietrukowicz} P.,  et~al., 2013, \actaa, \href
  {https://ui.adsabs.harvard.edu/abs/2013AcA....63..379P} {63, 379}

\bibitem[\protect\citeauthoryear{{Rattenbury}}{{Rattenbury}}{2009}]{Rattenbury2009}
{Rattenbury} N.~J.,  2009, \mn@doi [\mnras] {10.1111/j.1365-2966.2008.14074.x},
  \href {https://ui.adsabs.harvard.edu/abs/2009MNRAS.392..439R} {392, 439}

\bibitem[\protect\citeauthoryear{{Rattenbury} et~al.,}{{Rattenbury}
  et~al.}{2005}]{Rattenbury2005}
{Rattenbury} N.~J.,  et~al., 2005, \mn@doi [\aap] {10.1051/0004-6361:20052858},
  \href {https://ui.adsabs.harvard.edu/abs/2005A&A...439..645R} {439, 645}

\bibitem[\protect\citeauthoryear{{Ritter}}{{Ritter}}{1881}]{Ritter1881}
{Ritter} A.,  1881, \mn@doi [Annalen der Physik] {10.1002/andp.18812501207},
  \href {https://ui.adsabs.harvard.edu/abs/1881AnP...250..610R} {250, 610}

\bibitem[\protect\citeauthoryear{{Sackett}}{{Sackett}}{2001}]{sackett2001}
{Sackett} P.~D.,  2001, {Microlensing and the Physics of Stellar Atmospheres}.
p.~213

\bibitem[\protect\citeauthoryear{{Sajadian}}{{Sajadian}}{2015}]{sajadian2015}
{Sajadian} S.,  2015, \mn@doi [\mnras] {10.1093/mnras/stv1349}, \href
  {https://ui.adsabs.harvard.edu/abs/2015MNRAS.452.2587S} {452, 2587}

\bibitem[\protect\citeauthoryear{{Sajadian}}{{Sajadian}}{2016}]{sajadian2016}
{Sajadian} S.,  2016, \mn@doi [\apj] {10.3847/0004-637X/825/2/152}, \href
  {https://ui.adsabs.harvard.edu/abs/2016ApJ...825..152S} {825, 152}

\bibitem[\protect\citeauthoryear{{Sajadian} \& {Ignace}}{{Sajadian} \&
  {Ignace}}{2020}]{PaperI}
{Sajadian} S.,  {Ignace} R.,  2020, \mn@doi [\mnras] {10.1093/mnras/staa837},
  \href {https://ui.adsabs.harvard.edu/abs/2020MNRAS.494.1735S} {494, 1735}

\bibitem[\protect\citeauthoryear{{Sajadian} \& {Poleski}}{{Sajadian} \&
  {Poleski}}{2019}]{sajadian2019}
{Sajadian} S.,  {Poleski} R.,  2019, \mn@doi [\apj] {10.3847/1538-4357/aafa1d},
  \href {http://adsabs.harvard.edu/abs/2019ApJ...871..205S} {871, 205}

\bibitem[\protect\citeauthoryear{{Sako} et~al.,}{{Sako}
  et~al.}{2008}]{Sako2008}
{Sako} T.,  et~al., 2008, \mn@doi [Experimental Astronomy]
  {10.1007/s10686-007-9082-5}, \href
  {https://ui.adsabs.harvard.edu/abs/2008ExA....22...51S} {22, 51}

\bibitem[\protect\citeauthoryear{{Samus}, {Durlevich}  \& {Kazarovets}}{{Samus}
  et~al.}{1997}]{samus1997}
{Samus} N.~N.,  {Durlevich} O.~V.,   {Kazarovets} R.~V.,  1997, \mn@doi [Baltic
  Astronomy] {10.1515/astro-1997-0229}, \href
  {https://ui.adsabs.harvard.edu/abs/1997BaltA...6..296S} {6, 296}

\bibitem[\protect\citeauthoryear{{Samus}, {Kazarovets}, {Durlevich}, {Kireeva}
  \& {Pastukhova}}{{Samus} et~al.}{2017}]{samus2017}
{Samus} N.~N.,  {Kazarovets} E.~V.,  {Durlevich} O.~V.,  {Kireeva} N.~N.,
  {Pastukhova} E.~N.,  2017, \mn@doi [Astronomy Reports]
  {10.1134/S1063772917010085}, \href
  {https://ui.adsabs.harvard.edu/abs/2017ARep...61...80S} {61, 80}

\bibitem[\protect\citeauthoryear{{Schechter} \& {Wambsganss}}{{Schechter} \&
  {Wambsganss}}{2002}]{ray1}
{Schechter} P.~L.,  {Wambsganss} J.,  2002, \mn@doi [\apj] {10.1086/343856},
  \href {https://ui.adsabs.harvard.edu/abs/2002ApJ...580..685S} {580, 685}

\bibitem[\protect\citeauthoryear{{Schneider}, {Ehlers}  \& {Falco}}{{Schneider}
  et~al.}{1992}]{ray2}
{Schneider} P.,  {Ehlers} J.,   {Falco} E.~E.,  1992, {Gravitational Lenses},
  \mn@doi{10.1007/978-3-662-03758-4.
}

\bibitem[\protect\citeauthoryear{{Skowron} \& {Gould}}{{Skowron} \&
  {Gould}}{2012}]{Skowron2012}
{Skowron} J.,  {Gould} A.,  2012, arXiv[astro-ph.EP]: 1203.1034, \href
  {http://adsabs.harvard.edu/abs/2012arXiv1203.1034S} {}

\bibitem[\protect\citeauthoryear{{Soszy{\'n}ski}, {Udalski}, {Szyma{\'n}ski},
  {Kubiak}, {Pietrzy{\'n}ski}, {Wyrzykowski}, {Ulaczyk}  \&
  {Poleski}}{{Soszy{\'n}ski} et~al.}{2010}]{Igor2010}
{Soszy{\'n}ski} I.,  {Udalski} A.,  {Szyma{\'n}ski} M.~K.,  {Kubiak} J.,
  {Pietrzy{\'n}ski} G.,  {Wyrzykowski} {\L}.,  {Ulaczyk} K.,   {Poleski} R.,
  2010, \actaa, \href {https://ui.adsabs.harvard.edu/abs/2010AcA....60..165S}
  {60, 165}

\bibitem[\protect\citeauthoryear{{Soszy{\'n}ski} et~al.,}{{Soszy{\'n}ski}
  et~al.}{2013}]{Igor2013}
{Soszy{\'n}ski} I.,  et~al., 2013, \actaa, \href
  {https://ui.adsabs.harvard.edu/abs/2013AcA....63...21S} {63, 21}

\bibitem[\protect\citeauthoryear{{Soszy{\'n}ski} et~al.,}{{Soszy{\'n}ski}
  et~al.}{2014}]{Igor2014}
{Soszy{\'n}ski} I.,  et~al., 2014, \actaa, \href
  {https://ui.adsabs.harvard.edu/abs/2014AcA....64..177S} {64, 177}

\bibitem[\protect\citeauthoryear{{Sumi} et~al.,}{{Sumi}
  et~al.}{2013}]{sumi2013}
{Sumi} T.,  et~al., 2013, \mn@doi [\apj] {10.1088/0004-637X/778/2/150}, \href
  {https://ui.adsabs.harvard.edu/abs/2013ApJ...778..150S} {778, 150}

\bibitem[\protect\citeauthoryear{{Thomson}}{{Thomson}}{1862}]{Thompson1863}
{Thomson} W.,  1862, Proceedings of the Royal Society of London Series I, \href
  {https://ui.adsabs.harvard.edu/abs/1862RSPS...12..274T} {12, 274}

\bibitem[\protect\citeauthoryear{{Tisserand} et~al.,}{{Tisserand}
  et~al.}{2007}]{tisserand2007}
{Tisserand} P.,  et~al., 2007, \mn@doi [\aap] {10.1051/0004-6361:20066017},
  \href {https://ui.adsabs.harvard.edu/abs/2007A&A...469..387T} {469, 387}

\bibitem[\protect\citeauthoryear{{Valls-Gabaud}}{{Valls-Gabaud}}{1998}]{1998david}
{Valls-Gabaud} D.,  1998, \mn@doi [\mnras] {10.1046/j.1365-8711.1998.01247.x},
  \href {https://ui.adsabs.harvard.edu/abs/1998MNRAS.294..747V} {294, 747}

\bibitem[\protect\citeauthoryear{{Wyrzykowski}, {Udalski}, {Mao}, {Kubiak},
  {Szymanski}, {Pietrzynski}, {Soszynski}  \& {Szewczyk}}{{Wyrzykowski}
  et~al.}{2006}]{Lukaz2006}
{Wyrzykowski} L.,  {Udalski} A.,  {Mao} S.,  {Kubiak} M.,  {Szymanski} M.~K.,
  {Pietrzynski} G.,  {Soszynski} I.,   {Szewczyk} O.,  2006, \actaa, \href
  {https://ui.adsabs.harvard.edu/abs/2006AcA....56..145W} {56, 145}

\bibitem[\protect\citeauthoryear{{Wyrzykowski} et~al.,}{{Wyrzykowski}
  et~al.}{2015}]{OGLEIII}
{Wyrzykowski} {\L}.,  et~al., 2015, \mn@doi [\apjs]
  {10.1088/0067-0049/216/1/12}, \href
  {http://adsabs.harvard.edu/abs/2015ApJS..216...12W} {216, 12}

\bibitem[\protect\citeauthoryear{{Zheng} \& {M{\'e}nard}}{{Zheng} \&
  {M{\'e}nard}}{2005}]{zheng2005}
{Zheng} Z.,  {M{\'e}nard} B.,  2005, \mn@doi [\apj] {10.1086/497394}, \href
  {https://ui.adsabs.harvard.edu/abs/2005ApJ...635..599Z} {635, 599}

\makeatother
\end{thebibliography}
\end{document}